\documentclass[twocolumn]{aastex61}
\shortauthors{Brito \& Lopes}

\begin{document}

\title{A Theoretical Study of the Outer Layers of Eight {\it{Kepler}} F-stars: The Relevance of Ionization Processes \\ 
	}

\correspondingauthor{Ana Brito}
\email{ana.brito@tecnico.ulisboa.pt}

\author{Ana Brito}
\affil{Centro Multidisciplinar de Astrof\'isica\\ Instituto Superior T\'ecnico, Universidade T\'ecnica de Lisboa\\Av. Rovisco Pais, 1049-001 Lisboa, Portugal}
\affil{Departamento de Matem\'atica, Instituto Superior de Gest\~ao\\ Av. Marechal Craveiro Lopes, 1700-284, Lisboa, Portugal}

\author{Il\'idio Lopes}
\affiliation{Centro Multidisciplinar de Astrof\'isica\\ Instituto Superior T\'ecnico, Universidade T\'ecnica de Lisboa\\Av. Rovisco Pais, 1049-001 Lisboa, Portugal}

\begin{abstract}

We have analyzed the theoretical model envelopes of eight {\it{Kepler}} F-stars by computing the phase shift of the acoustic waves, $\alpha(\omega)$, and its related function, $\beta(\omega)$. The latter is shown to be a powerful probe of the external stellar layers since it is particularly sensitive to the partial ionization zones located in these upper layers. We found that these theoretical envelopes can be organized into two groups, each of which is characterized by a distinct $\beta(\omega)$ shape that we show to reflect the differences related to the magnitudes of ionization processes. Since $\beta(\omega)$ can also be determined from the experimental frequencies, we compared our theoretical results with the observable $\beta(\omega)$. 
Using the function $\beta(\omega)$, and with the purpose of quantifying the magnitude of the ionization processes occurring in the outer layers of these stars, we define two indexes, $\Delta \beta_1$ and $\Delta \beta_2$. These indexes allow us to connect the microphysics of the interior of the star with macroscopic observable characteristics.
Motivated by the distinct magnetic activity behaviors of F-stars, we studied the relation between the star's rotation period and these indexes. We found a trend, in the form of a power-law dependence, that favors the idea that ionization is acting as an underlying mechanism, which is crucial for understanding the relation between rotation and magnetism and even observational features such as the Kraft break.

\end{abstract}

\keywords{{stars: activity -- stars: general -- stars: interiors -- stars: oscillations -- stars: rotation -- stars: solar-type}}

\section{Introduction} \label{sec:intro}

\begin{figure*}
	\plottwo{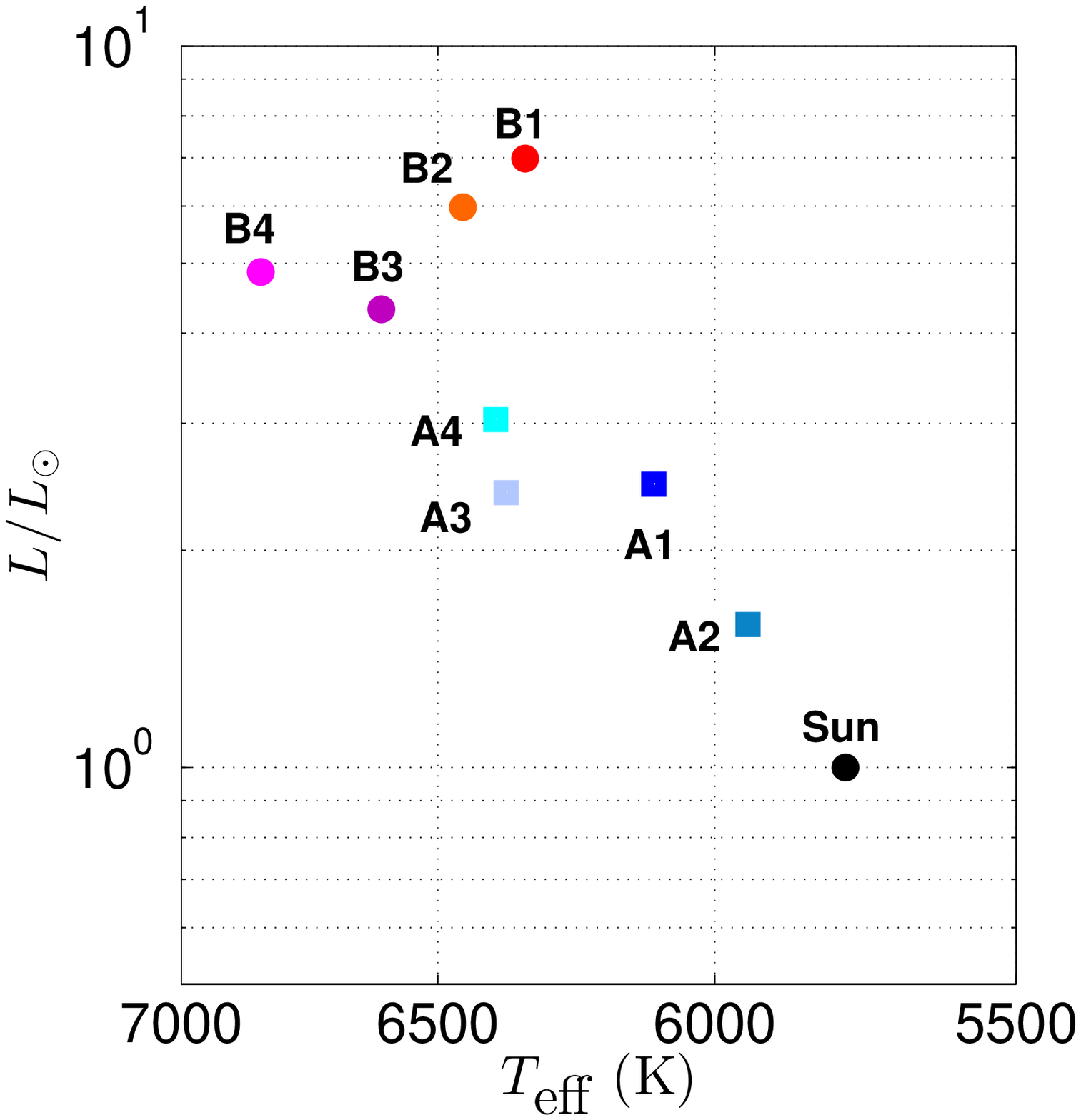}{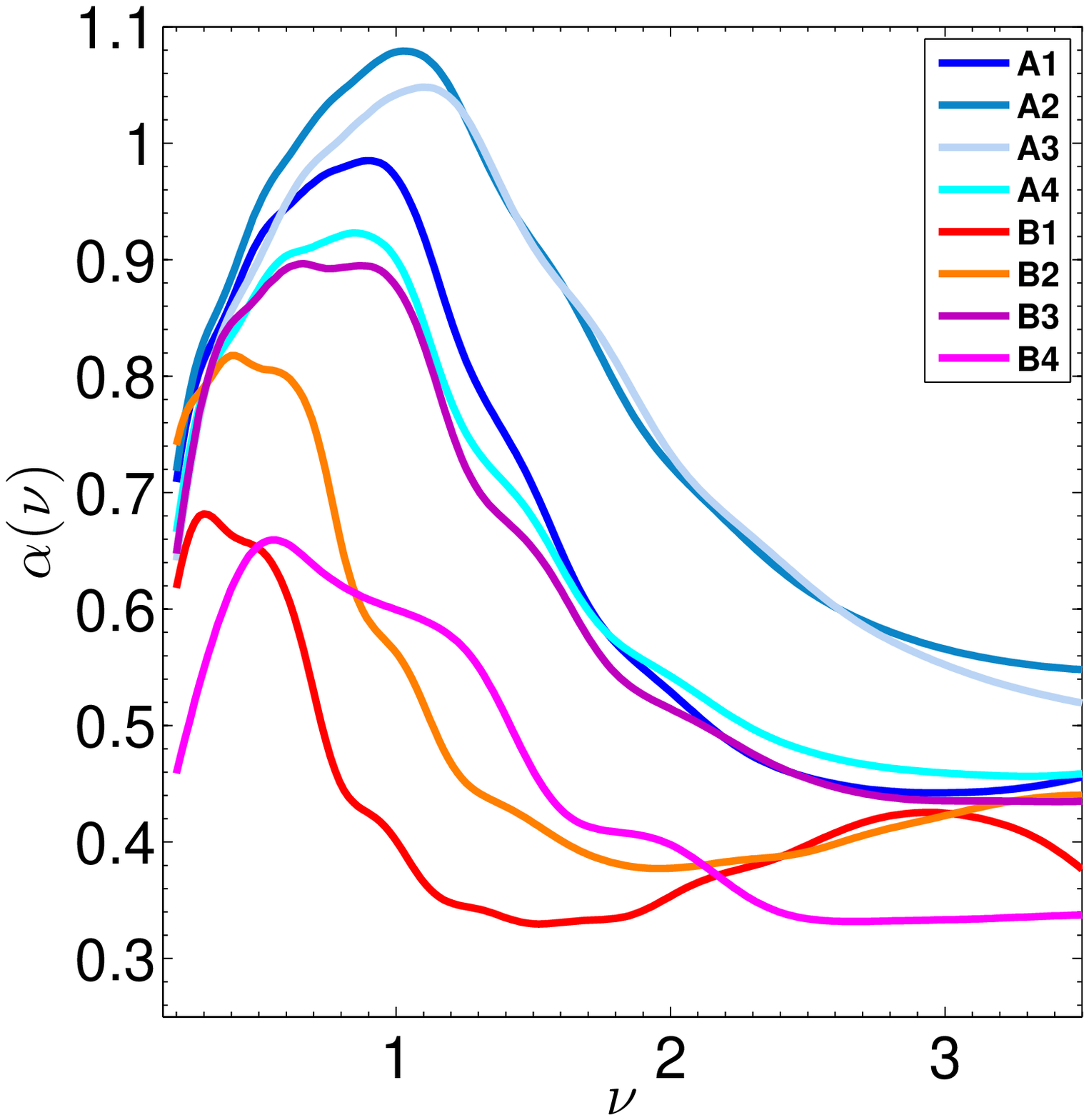}
	\caption{ Left: theoretical H--R diagram with luminosities plotted as a function of effective temperature. The squares represent the group A of models, the circles represent the B group, and the black marker stands for the Sun. Right: the phase $\alpha$, obtained numerically from equation (\ref{eq:(9)}), as a function of the cyclic frequency $\nu$ ($\nu=\omega/{2\pi}$).\label{fig:fig1}}
\end{figure*}

\begin{deluxetable*}{ccCclcc}
	\tablecaption{Seismic and non-seismic observational constraints of the selected targets \label{table:1}}
	\tablecolumns{7}
	\tablenum{1}
	\tablewidth{0pt}
	\tablehead{
		\colhead{Star Id. (KIC) } &
		\colhead{MK} &
		\colhead{$\langle \Delta \nu \rangle$ ($\mu $Hz) } & 
		\colhead{$\nu_{\text{max}}$ ($\mu $Hz) } & 
		\colhead{$T_{\text{eff}}$ (K) } &
		\colhead{log $g$ } &
		\colhead{[Fe/H] (dex)}
	}
	\startdata
	A1 (8228742)  & F9IV-V & $ 62.071^{+0.022}_{-0.021}$ & $ 1190.5^{+3.4}_{-3.7}$ & 6122 $\pm$ 77 & $ 4.03^{+0.004}_{-0.005}$ & -0.08 $\pm$ 0.10\\
	A2 (6116048)  & F9IV-V & $100.754^{+0.017}_{-0.017}$ & $2126.9^{+5.5}_{-5.0}$ & 5895 $\pm$ 70 & $4.19^{+0.08}_{-0.08}$ &-0.26 $\pm$ 0.07\\
	A3 (10454113) & F9IV-V & $105.063^{+0.031}_{-0.033}$  & $2357.2^{+8.2}_{-9.1}$ & 6216 $\pm$ 68 & $4.46^{+0.10}_{-0.10}$ & 0.00 $\pm$ 0.05\\
	A4 (12009504) & F9IV-V  & $88.217^{+0.026}_{-0.025}$ & $1865.6^{+7.7}_{-6.2}$ & 6267 $\pm$ 71 & $4.37^{+0.11}_{-0.11}$ &-0.03 $\pm$ 0.06\\
	\hline
	B1 (6679371)  & F5IV-V  & $50.601^{+0.029}_{-0.029}$ & $941.8^{+5.1}_{-5.0}$ & 6344 $\pm$ 131& $3.92^{+0.21}_{-0.21}$ &-0.10 $\pm$ 0.21\\
	B2 (7103006)  & F8IV    & $59.658^{+0.029}_{-0.030}$ & $1167.9^{+7.2}_{-6.9}$ & 6485 $\pm$ 86 & $4.50^{+0.11}_{-0.11}$ & 0.19 $\pm$ 0.06\\
	B3 (9139163)  & F8IV   & $81.170^{+0.042}_{-0.036}$ & $1729.8^{+6.2}_{-5.9}$ & 6577 $\pm$ 69 & $4.44^{+0.10}_{-0.10}$ & 0.21 $\pm$ 0.06\\
	B4 (9206432)  & F8IV  & $84.926^{+0.046}_{-0.051}$ & $1866.4^{+10.3}_{-14.9}$ & 6772 $\pm$ 73 & $4.61^{+0.11}_{-0.11}$ & 0.28 $\pm$ 0.06\\
	\enddata
\end{deluxetable*}

\begin{deluxetable*}{ccCcccccc} 
	\tablecaption{Parameters of the optimal models obtained for the stars of the sample with the code CESAM. \label{table:2}}
	\tablecolumns{9}
	\tablenum{2}
	\tablewidth{0pt}
	\tablehead{
		\colhead{Star Id.} &
		\colhead{${M}/{M}_\odot$} &
		\colhead{${R}/{R}_\odot$} & 
		\colhead{${L}/{L}_\odot$} & 
		\colhead{$T_{\text{eff}}$ (K)} &
		\colhead{Age (Gyr)} &
		\colhead{$Y_0$} &
		\colhead{$\alpha$} &
		\colhead{$r_{\text{bcz}}/R$}
	}
	\startdata 
	A1 & 1.24    & 1.409  & 2.470 & 6107    & 3.802 & 0.248 & 1.65 & 0.803 \\ 
	A2 & 0.99    &1.187   & 1.579  & 5943   & 7.259 & 0.275 & 1.70 & 0.740 \\
	A3 & 1.20    &1.271   & 2.404  & 6383   & 3.782 & 0.253 & 2.30 & 0.775 \\
	A4 & 1.20    &1.424   & 3.036  & 6392   & 2.982 & 0.287 & 1.80 & 0.847 \\
	\hline
	B1 & 1.55    & 2.195 & 6.984  & 6338  & 2.182 & 0.249 & 1.50 & 0.929  \\
	B2 & 1.45    & 1.960 & 5.979  & 6453  & 1.982 & 0.313 & 1.90 & 0.871  \\
	B3 & 1.40    & 1.589 & 4.318  & 6607  & 1.582 & 0.308 & 2.00 & 0.879  \\
	B4 & 1.50    & 1.572 & 6.741  & 6741  & 0.472 & 0.326 & 1.80 & 0.951  \\
	\enddata
\end{deluxetable*}

Recent advances in asteroseismology have been exceptional primarily due to space-based missions. With its high quality data sets, the {\it{Kepler}} mission has revolutionized the studies in this field \citep{2010PASP..122..131G, 2011Sci...332..213C}. In addition to the exceptional quality of the data, the quantity of data, now available in unprecedented numbers, is leading to new breakthroughs  in stellar structure, stellar dynamics, and stellar evolution. The launch of the {\it{TESS}} mission \citep{2015JATIS...1a4003R} scheduled to take place within the next year, and also, in the not so far future, the launch of the {\it{PLATO}} mission \citep{2014ExA....38..249R}, should improve even further the quality and quantity of seismological data. These missions will provide oscillation frequencies for a large number of targets in many directions of the galaxy. Ensemble asteroseismology, the asteroseismology inside clusters of stars, allows for the probing scaling laws, for studying stellar activity, and even for testing theories of stellar evolution \citep{2011Sci...332..213C, 2013ARA&A..51..353C}.

However, uncertainties in stellar physics exist, and they have an adverse impact on these very same theories. These uncertainties are exacerbated if we aim to give a physical description of the outermost layers of solar-type stars. Outer layers are poorly described by current stellar models.
In these layers we have an intricate web of different processes occurring and interacting in the same region and on the same time scale: convection, magnetic fields, and pulsations. Beneath the stellar surface (a few percent of the star's radius) this interaction becomes even more complicated due to the partial ionization processes of chemical elements.
All this activity contributes to a complex background structure that affects the frequencies of stellar oscillations in several ways. Furthermore, solar-like oscillations are excited and damped by the near-surface convection.

It is therefore extremely important to probe the physical processes that are taking place in these more external layers of stars. These  processes can be inferred by an appropriate seismic diagnostic like the $\alpha(\omega)$ -- acoustic phase shift.  This quantity measures the phase shift experienced by acoustic waves in the surface layers of stars. Although it  is known to be a  reliable observable for obtaining information about the physics of the external regions of the Sun~\citep{1989SvAL...15...27B, 1997ApJ...480..794L, 2001MNRAS.322..473L, 2001MNRAS.322...85R}, here we apply the technique to other stars. In this paper we analyze and discuss the signature of this phase and its dependence on frequency, $\alpha(\omega)$, for a group of eight Kepler solar-type stars. We also use, as a seismic probe of the external stellar layers, the derivative of $\alpha(\omega)$. This is the seismic parameter $\beta(\omega)$,  which can be computed from the frequencies of low-degree modes \citep{1987SvAL...13..179B, 1989ASPRv...7....1V, 1997ApJ...480..794L,2001MNRAS.322..473L}.  In the past this diagnostic method has yielded important results for the Sun. Namely, it enabled the measurement of the solar helium abundance by \citet{1991Natur.349...49V}, and also contributed to the calibration of the equation of state \citep{1991A&A...248..263P, 1992MNRAS.257...32V}. It is a powerful diagnostic tool, particularly suited to studying the external layers of solar-like stars.

For the eight selected stars, we have computed the theoretical dependence of the phase shift on the frequency, $\alpha(\omega)$. We also calculated the seismic observable $\beta(\omega)$ and compared it with the theoretical seismic parameter, $\beta(\omega)$. The theoretical $\beta(\omega)$ was obtained via two different processes: from the structural parameters of the envelopes of stellar models, and also from theoretical tables of frequencies. We found an overall good agreement between the theoretical $\beta(\omega)$ and the observational $\beta(\omega)$ for all the stars, particularly if we consider a frequency interval in the vicinity of the observed value of the frequency of maximum power. However, the theoretical seismic signatures reveal several distinct oscillatory behaviors that we know are related to the partial ionization processes. These behaviors allow us to split our set of eight models into two subsets of four models each, which we will designate subsets A and B. The subset A has a $\beta(\omega)$ signature that is more "Sun-like", whereas the subset B exhibits a $\beta(\omega)$ with characteristics somewhat different from the shape of a "sun-like" $\beta(\omega)$. The distinction is related to the intensity and location of ionization processes, which in turn are reflected in the oscillatory character of $\beta(\omega)$. 

Stars on the main-sequence are known to follow two distinct rotational regimes \citep[e.g.,][]{2013ApJ...776...67V}. These rotational regimes are mainly determined by the effectiveness with which stars can lose their angular momentum from birth. Cool stars with $T_{\text{eff}}<6200$ K rotate slowly, with rotation periods larger than 10 days. Because of their thick convective envelopes, they lose angular momentum due to the presence of magnetic winds in their atmospheres, quickly  losing their initial rotational conditions. On the other hand, hot stars, with $T_{\text{eff}}>6200$ K, are rapid rotators. These stars are not efficient in the generation of magnetic winds because their convective envelopes are not thick enough. Hence, they do not lose significant amounts of angular momentum and remain rapidly rotating from birth. The sharp transition from slowly rotating main-sequence stars to rapidly rotating main-sequence stars is known as the Kraft break and occurs around approximately at $1.3 \, M_{\odot}$ \citep{1967ApJ...150..551K}.
Stellar rotation period and the stellar activity cycle are also related \citep[e.g.,][]{1984ApJ...287..769N, 1996ApJ...460..848B, 2003A&A...397..147P, 2007ApJ...657..486B, 2011ApJ...743...48W}. The trend can be summed up in the formula $P_{cyc} \propto P^q_{rot}$ with $q=1.25 \pm 0.5$ \citep{1984ApJ...287..769N}. A follow-up analysis revealed two related branches activity-wise: fast rotators with short  cycles and slow rotators with long cycles. Rotators with intermediate rates of rotation can display both kinds of magnetic behavior \citep{1999ApJ...524..295S, 2002ASPC..277..311S}. Based on these results, \citet{2007ApJ...657..486B} suggested that the two branches illustrate the action of two dynamos in different regions of a star.
In our chosen sample of eight Kepler stars, the two groups of stars with distinct rotation periods can be related to different ionization signatures occurring in the stellar envelopes.
The group of theoretical models that exhibits, in accordance with our diagnostic, a more intense pattern of ionization processes in the envelope, represents the group of stars that are known to have shorter rotation periods. Conversely, the group of theoretical models where the magnitude of ionization processes is diminished, represents the group of stars with larger rotation periods. Therefore, ionization can be a powerful instrument to extend our knowledge about the rotation of solar-type stars, and also to test evolutionary theories.

The paper is organized as follows. In Section 2 we describe the method: the astroseismological diagnostic $\beta(\omega)$. Section 3 introduces the targets, describes the models, and probes the envelopes of the theoretical models. The theoretical seismic parameter $\beta(\omega)$ is compared with the observable $\beta(\omega)$. In section 4 we introduce two indexes that measure the magnitude of ionization processes in the envelope and relate them with the rotation period of the star. Finally, Section 5 refers to the conclusions.

\begin{figure*}
	\centering
	\includegraphics[width=0.6\textwidth]{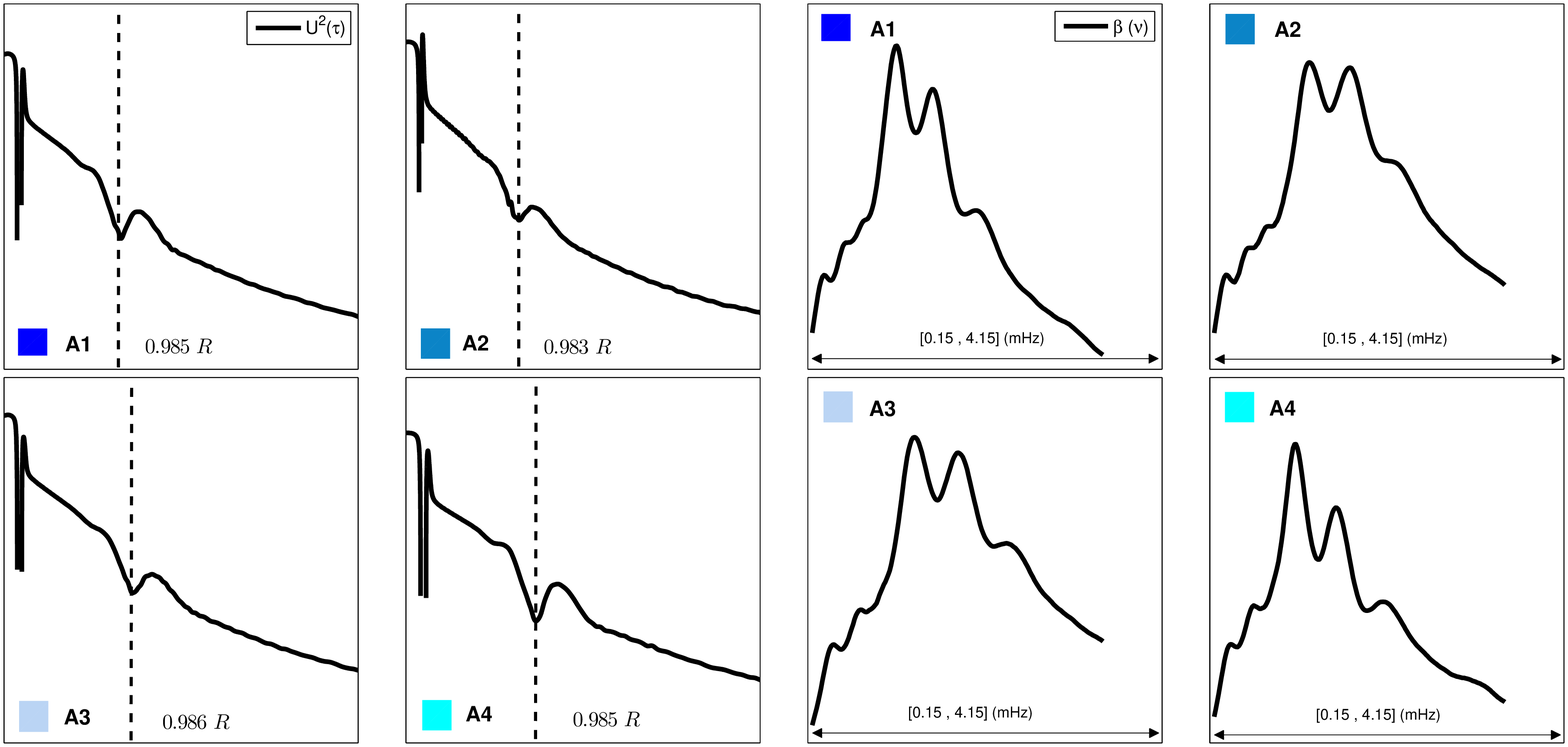}\\
	\includegraphics[width=0.6\textwidth]{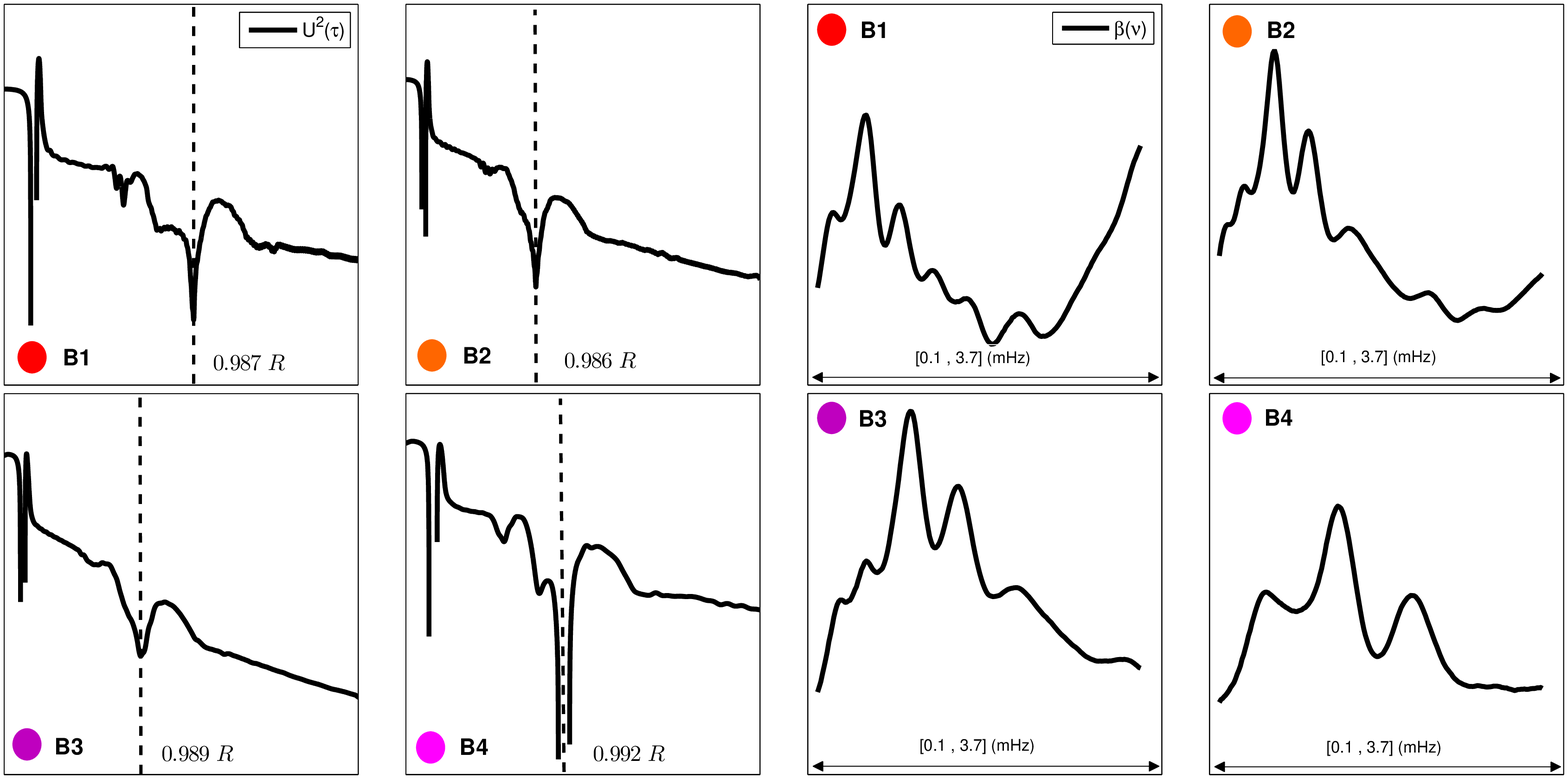}
	\caption{Detailed qualitative theoretical description of the upper layers of the eight Kepler stars between the photosphere  and the region above the base of the convective zone (BCZ). This description highlights ionization processes occurring in the envelope, with an emphasis on the partial ionization of light elements. Top left panel: acoustic potential computed from the envelope of stellar models for the group A of stars. Top right panel: theoretical seismic signatures $\beta(\nu)$ obtained from the corresponding acoustic potentials (group A). These signatures were obtained by numerical differentiation of $\alpha / \omega$, where $\omega = 2 \pi \nu$ represents the angular frequency of the mode. Bottom left panel: acoustic potential computed from the envelope of stellar models for the group B of stars. Bottom right panel: theoretical seismic signatures $\beta(\nu)$ obtained from the corresponding acoustic potentials (group B). These signatures were obtained by numerical differentiation of $\alpha / \omega$, where $\omega = 2 \pi \nu$ represents the angular frequency of the mode. \label{fig:fig2}}
\end{figure*}


\section{Propagation of acoustic waves in the atmospheres of stars}  \label{sec:2}

\begin{figure}
	\centering
	\includegraphics[width=0.4\textwidth]{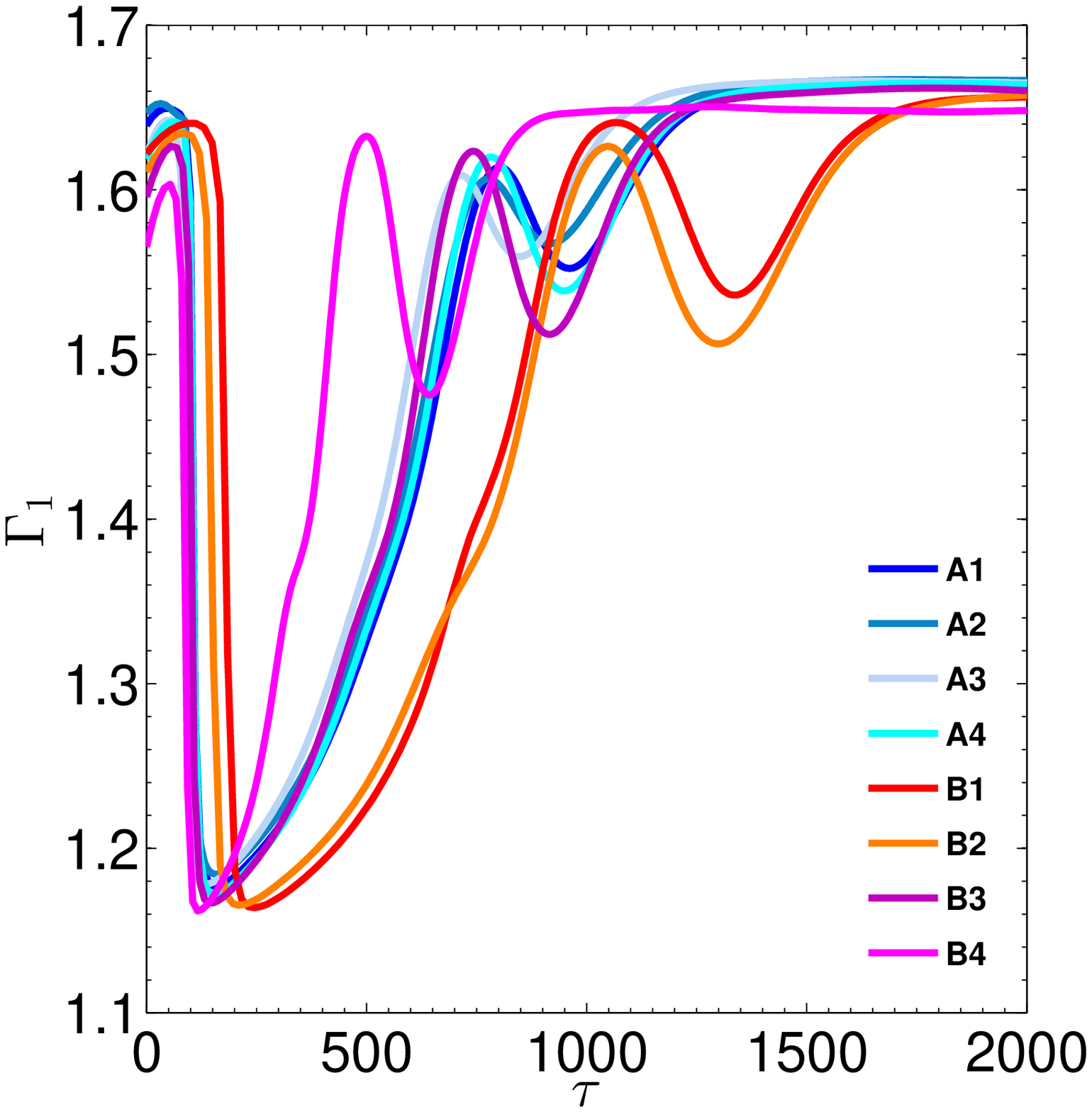}
	\caption{Adiabatic index, $\Gamma_1$, as a function of the acoustic depth $\tau$ for the eight stellar models. The stars of group A exhibit weaker ionization bumps than the stars in group B.\label{fig:fig3}}
\end{figure}

Oscillation frequencies, as is well-known, carry information about the internal regions of the stars. This information can be extracted and separated into two parts by an appropriate handling of the experimental data.
One part of the information is asymptotic in nature and reflects the structure of stellar interiors. The asymptotic theory of non-radial acoustic oscillations allows us to solve inverse problems such as the distribution of the sound speed, density, and Brunt--V\"{a}is\"{a}l\"{a} frequency in the interior of stars.
The second part of the information concerns the regions of the stars where the hypothesis of asymptotic analysis is no longer valid. Accordingly, this part of the information reflects the properties of the outer layers of stars. Indeed, in these layers the wavelength of acoustic waves is smaller or of the same order as the variation of the background state. These regions are responsible for partial wave reflection and hence for the establishment of a resonant cavity. The information carried by the trapped internal waves can be extracted from observational data and takes the form of a frequency dependence of the phase shift $\alpha$, which results from reflection. The outermost layers of a star are also the layers where a breakdown of the adiabaticity occurs and where we have inhomogeneities associated with convection and magnetic fields.

The equations describing the non-radial adiabatic acoustic oscillations in the stellar surface \citep[e.g.][]{1979nos..book.....U, 1984ARA&A..22..593D, 2010aste.book.....A} are obtained within two main approximations. One is the Cowling approximation, which neglects the Eulerian perturbation to the gravitational potential. The second takes into account the fact that for low-degree modes, all the trajectories of the sound waves trapped in the stellar interior are almost vertical. The radial component of the wavevector near the surface is much greater than the horizontal one, and this makes the oscillations to depend on the frequency alone. Hence, the outer phase shift is a function of frequency $\alpha = \alpha(\omega)$. This phase shift is then determined by the structure of the more external stellar layers where the reflection of internal acoustic waves occurs.

The problem created by the fact that an asymptotic description is invalid in the upper layers of the stars can be eliminated by matching the asymptotic solutions in the stellar interior with the exact solution near the surface. This procedure leads to an equation for the eigenfrequencies that contain a frequency dependence on the phase shift. The dispersion relation of the stationary acoustic waves is given by
\begin{equation} \label{eq:(1)}
F(w) = \frac{\pi}{\omega} \alpha(\omega) + \frac{\pi}{\omega} n \; ,
\end{equation}
where $w = \omega/L$ with $L=l+1/2$. Here, $\omega$ is the angular frequency and $n$ and $l$ are, respectively, the radial order and the degree of the mode.  $F(w)$ is determined by the radial distribution of the sound speed in the stellar interior, whereas $\alpha(\omega)$ stands for the total phase shift of the reflected by the surface acoustic waves. This expression is known as the Duvall law \citep{1982Natur.300..242D} and provides a bridge between theory and observational data.

\begin{figure*}
	\plottwo{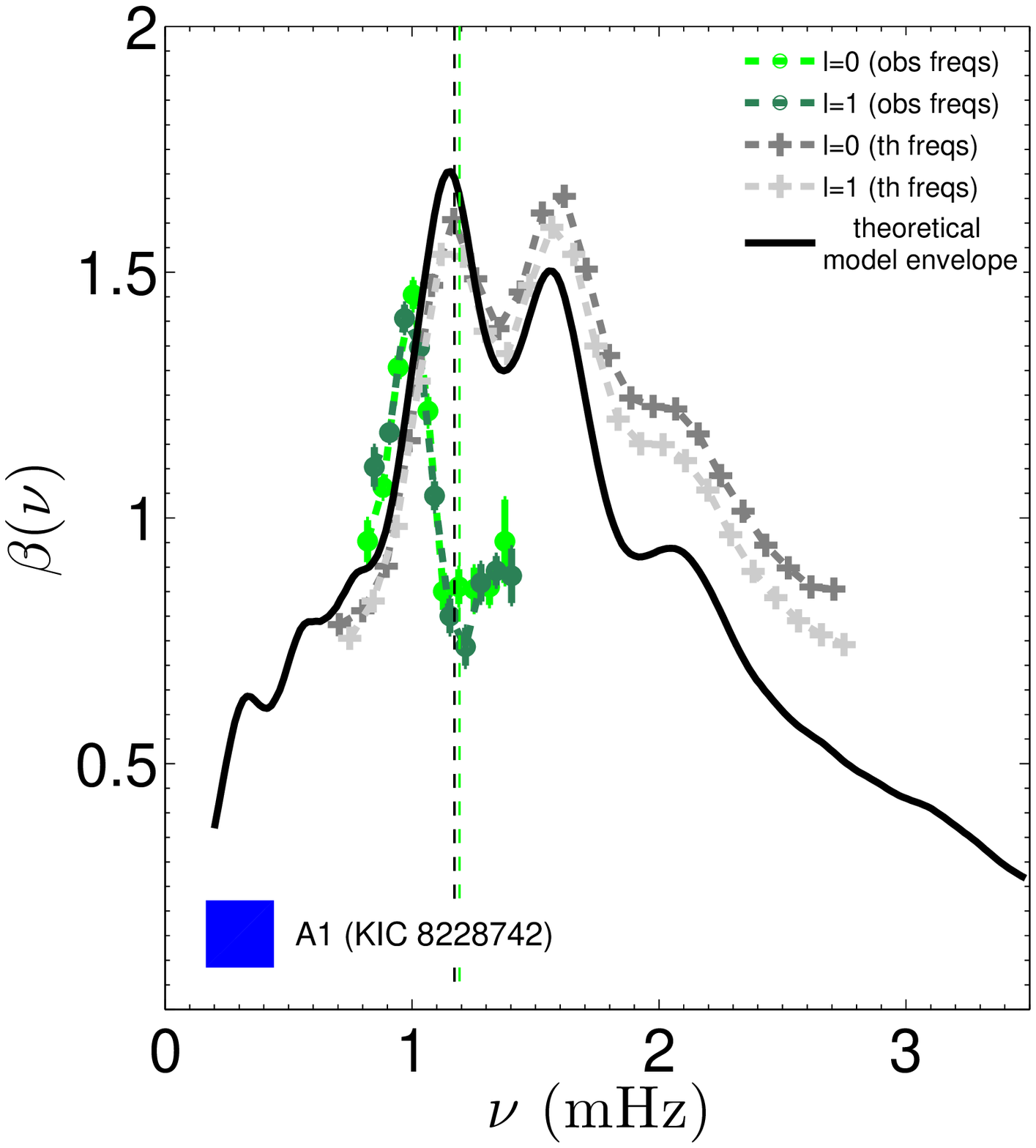}{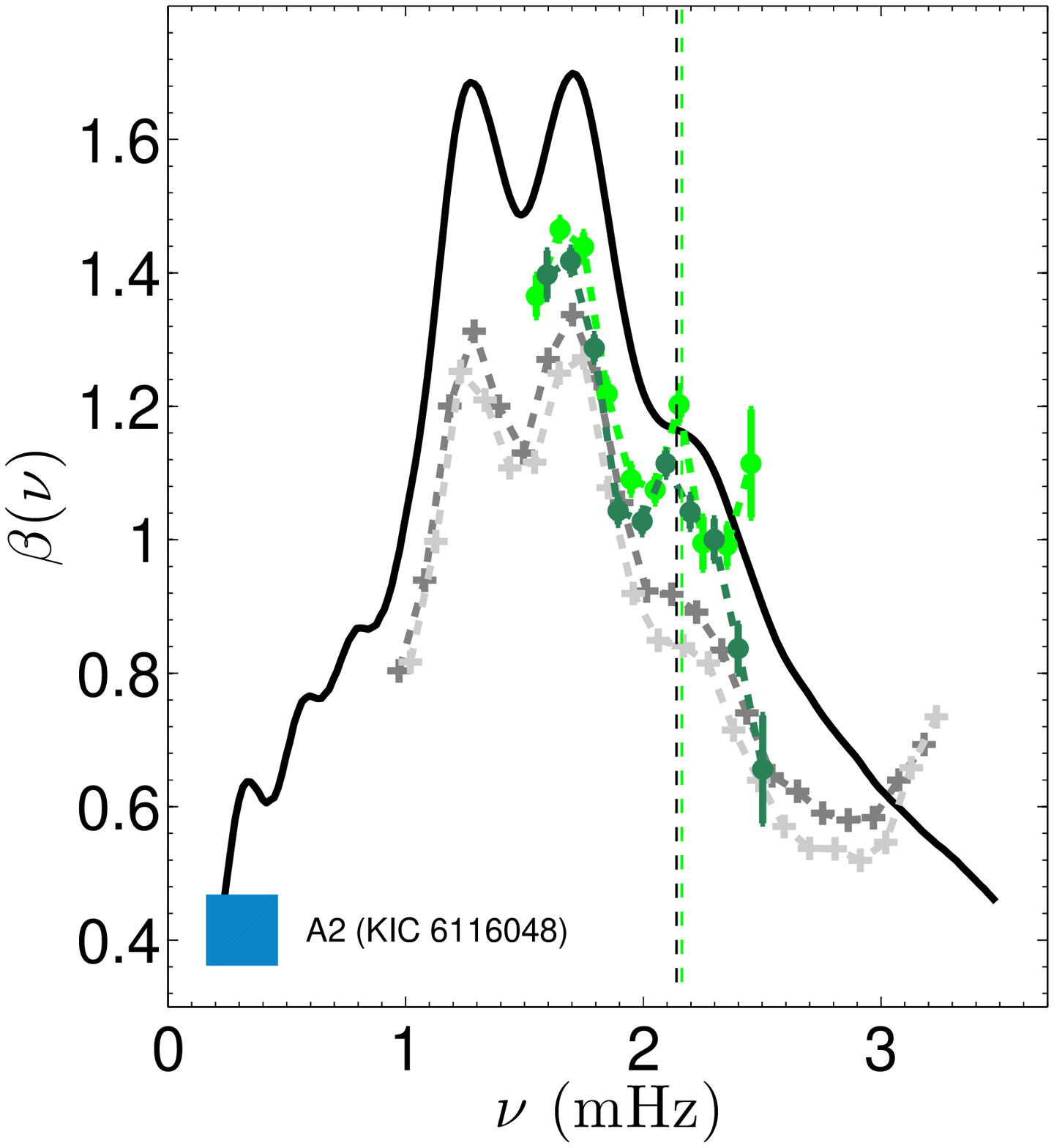}
	\plottwo{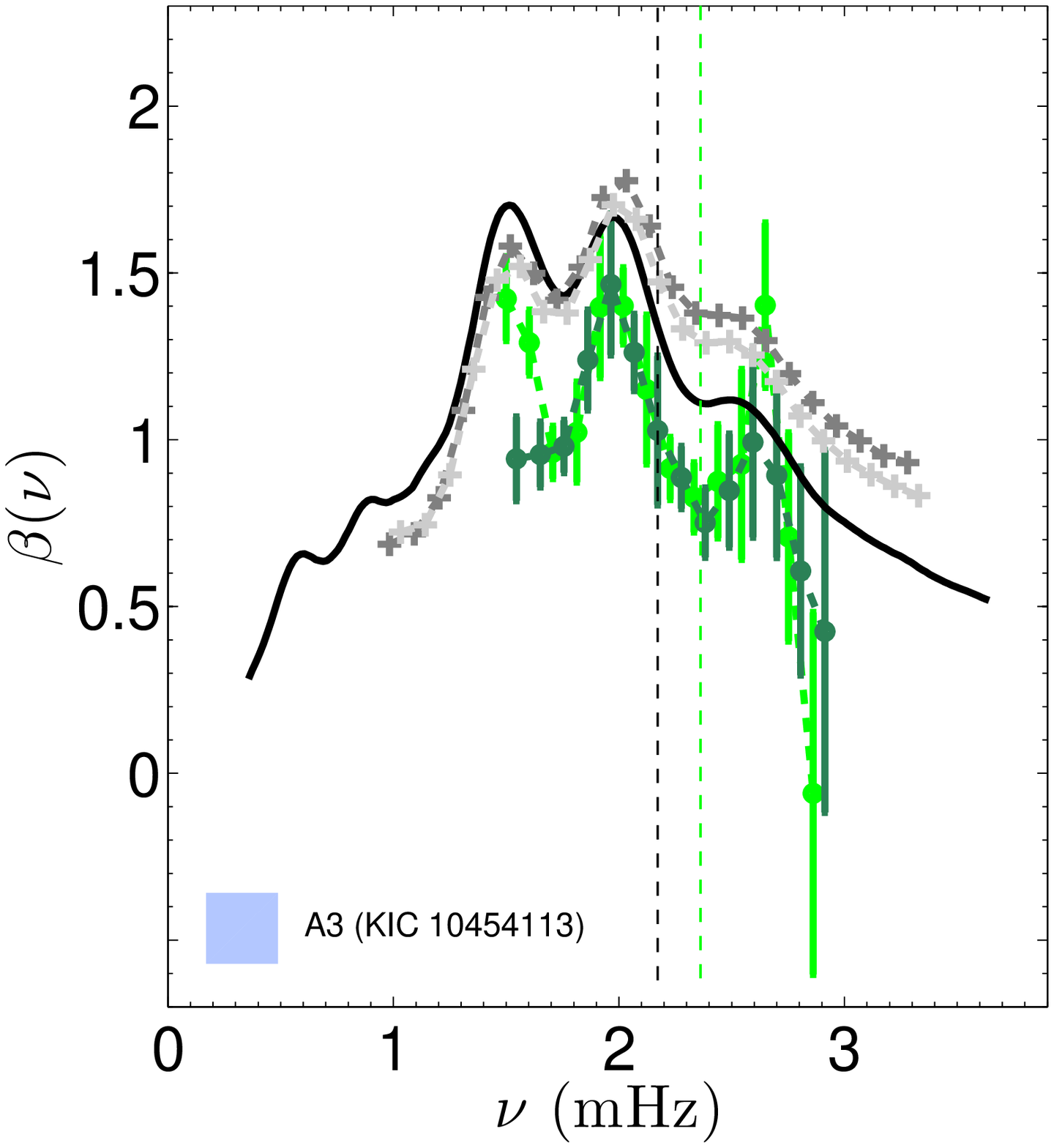}{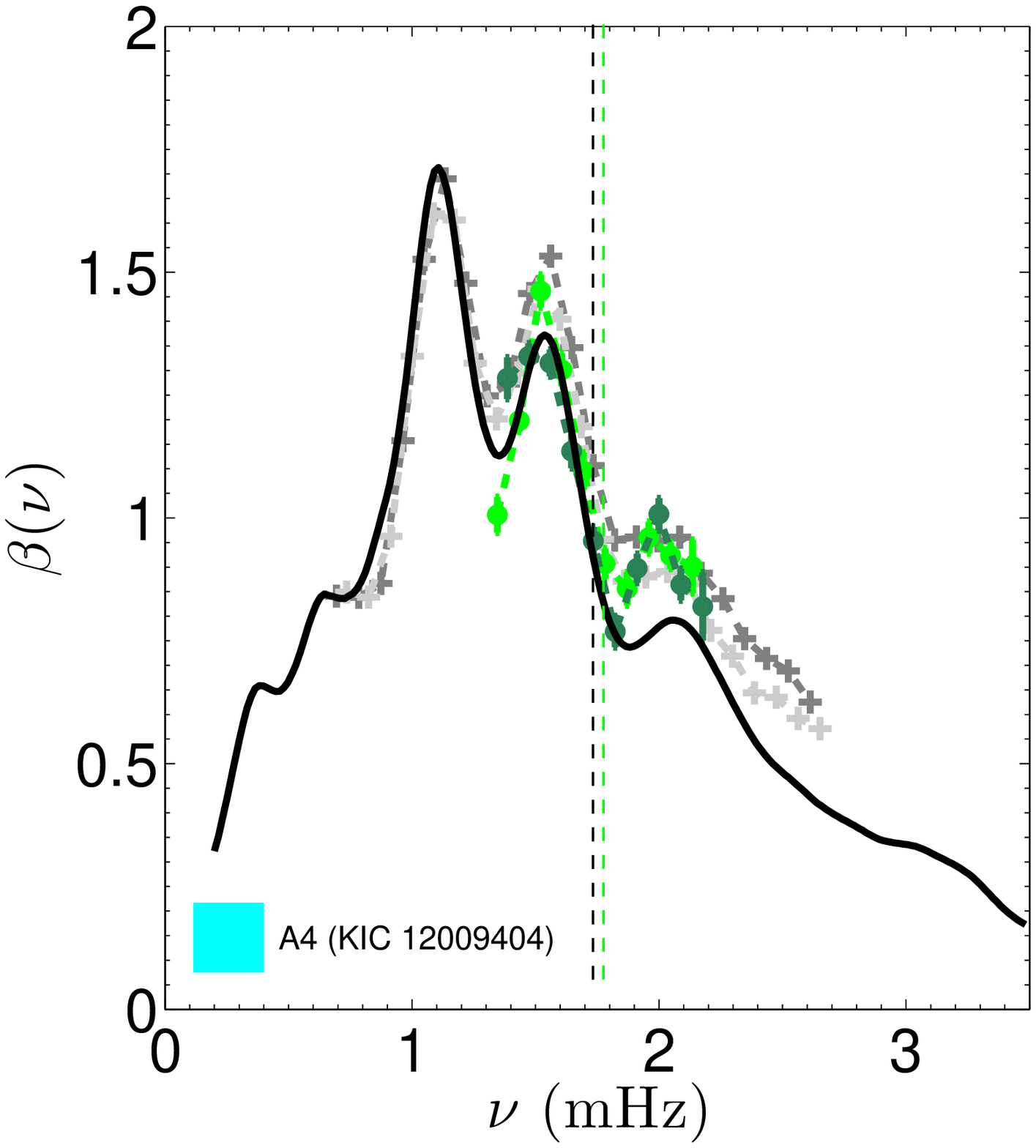}
	\caption{Group A. Black solid line: seismic parameter $\beta(\nu)$ computed from the theoretical model envelope. Gray symbols: theoretical $\beta(\nu)$ computed from a table of frequencies for the degrees $l=0$ (dark gray) and $l=1$ (light gray). Green dashed lines: seismic observable $\beta(\nu)$ computed from an observational frequency table for modes with degree $l = 0$ (light green) and $l=1$ (dark green). The vertical green dashed line indicates the approximate position of the observational frequency at maximum power $\nu_{\text{max}}$. The vertical black dashed line indicates the approximate position of the frequency at maximum power, $\nu_{\text{max}}$, obtained from the scaling relation in \citet{1995A&A...293...87K}. \label{fig:fig4}}
\end{figure*}

\begin{figure*}
	\plottwo{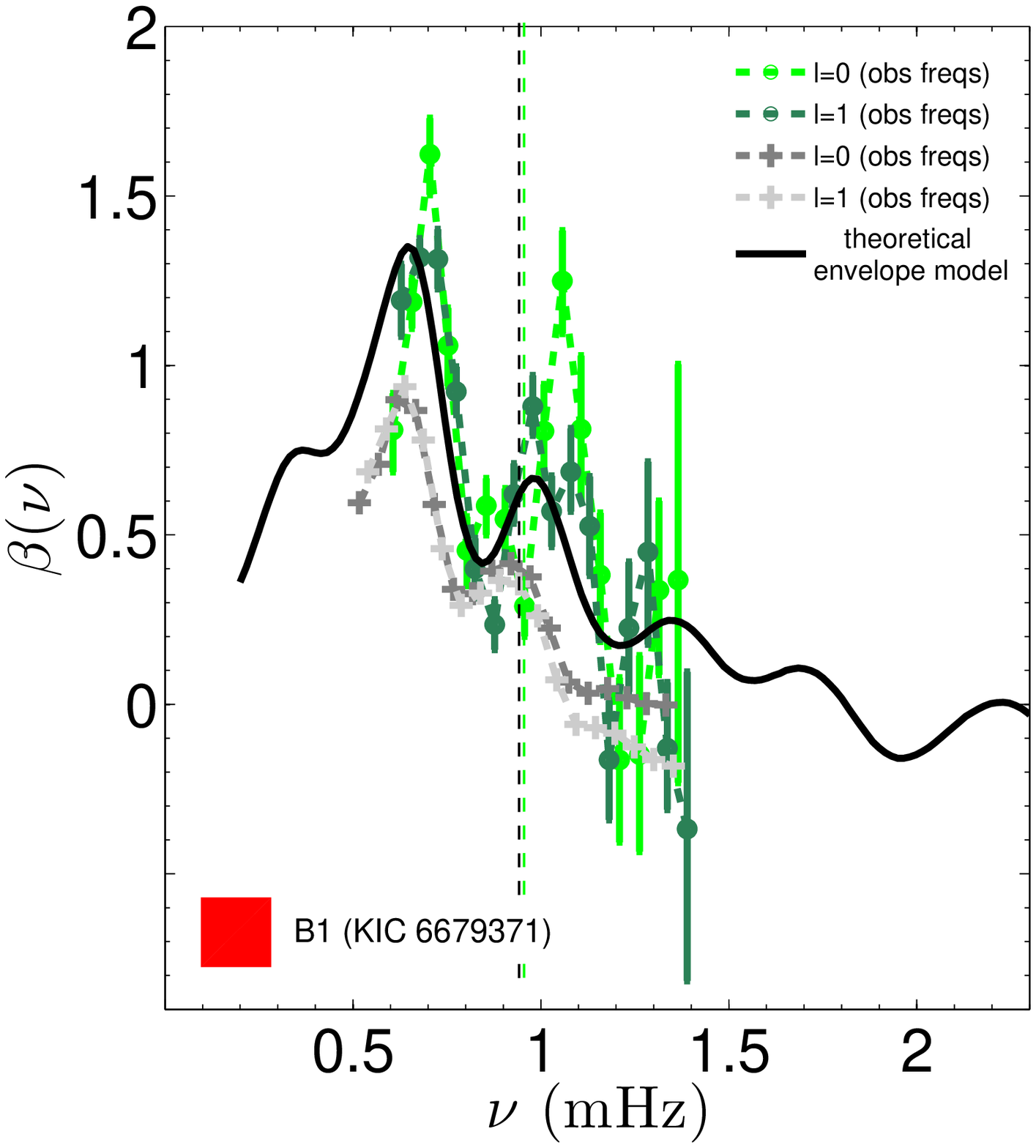}{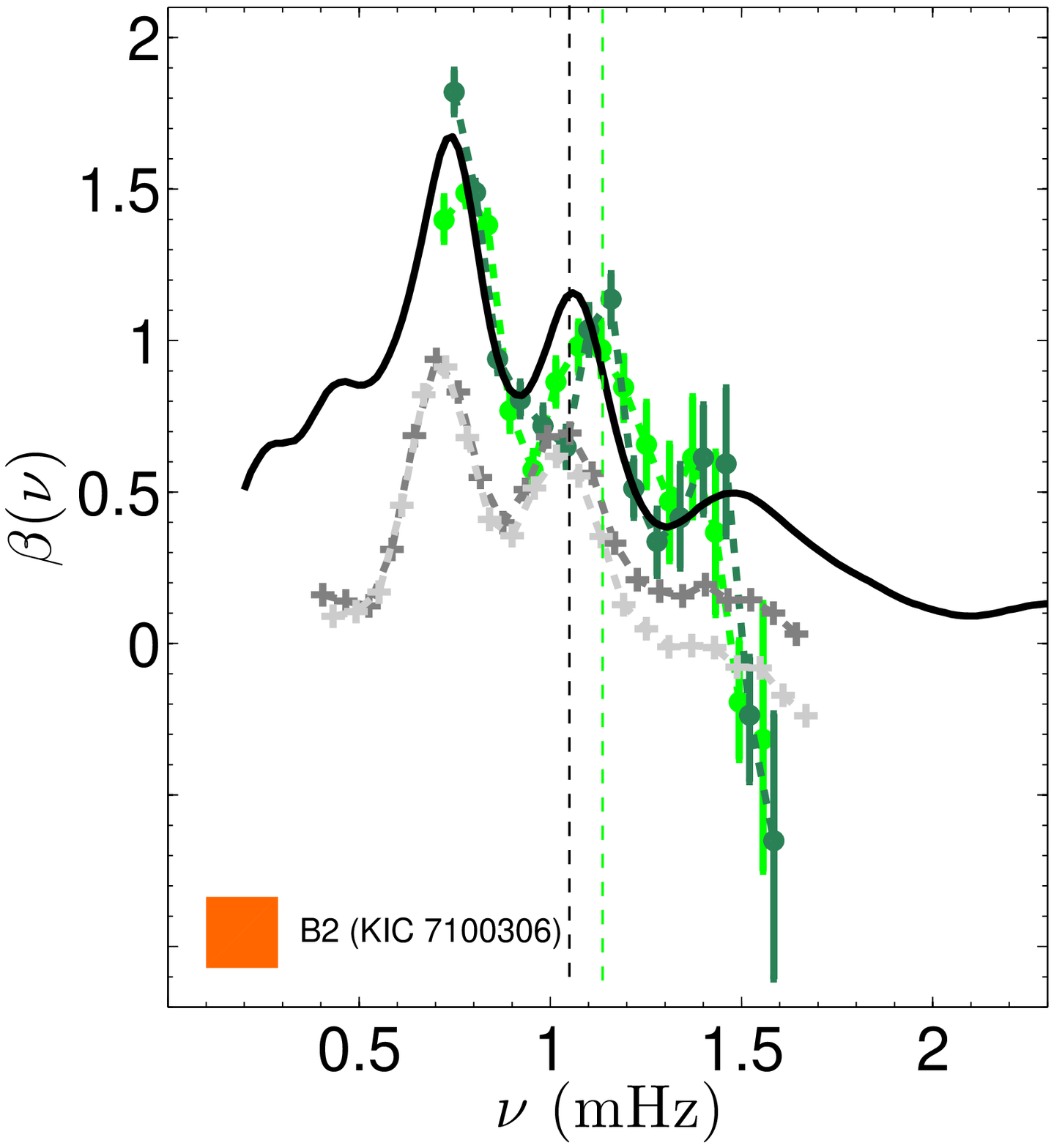}
	\plottwo{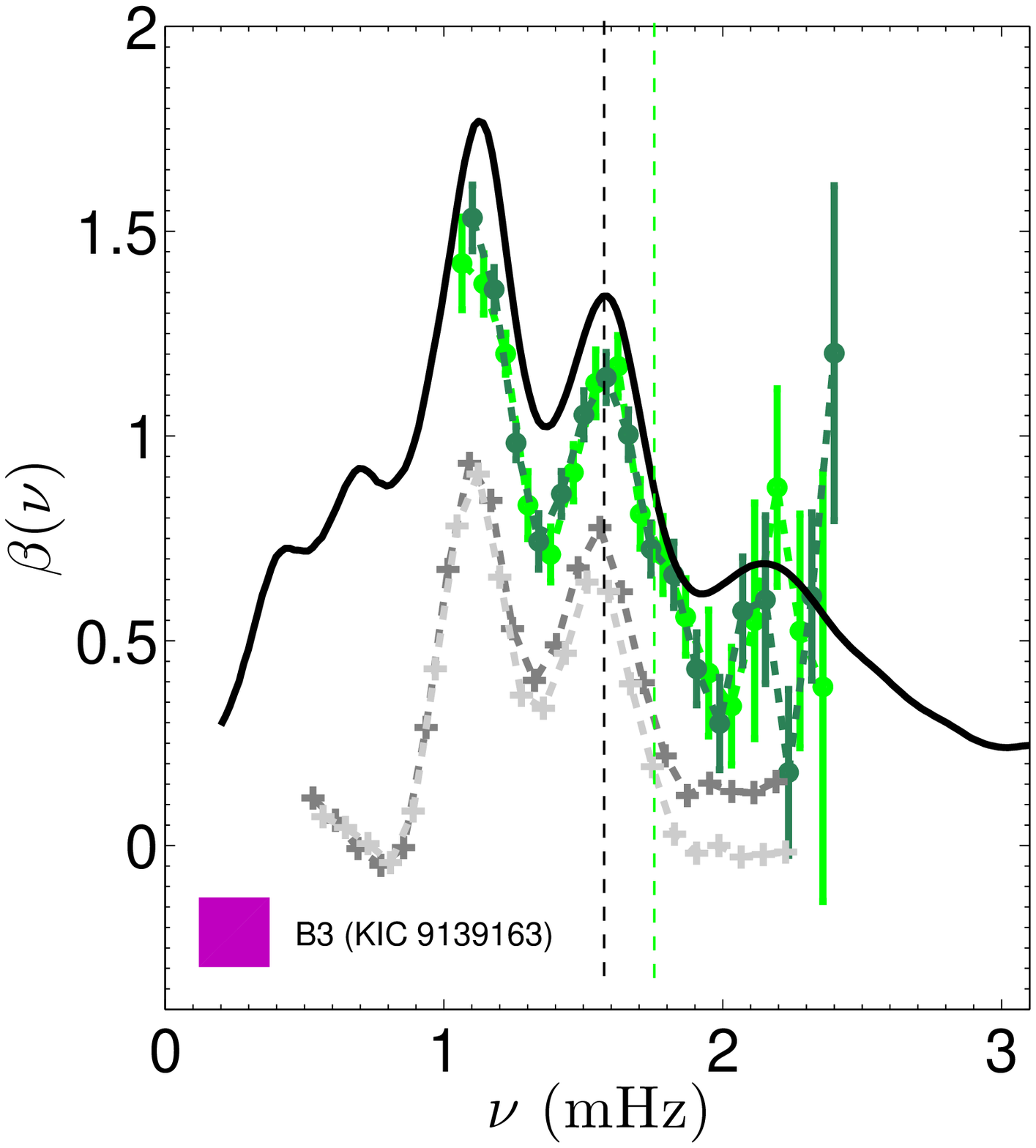}{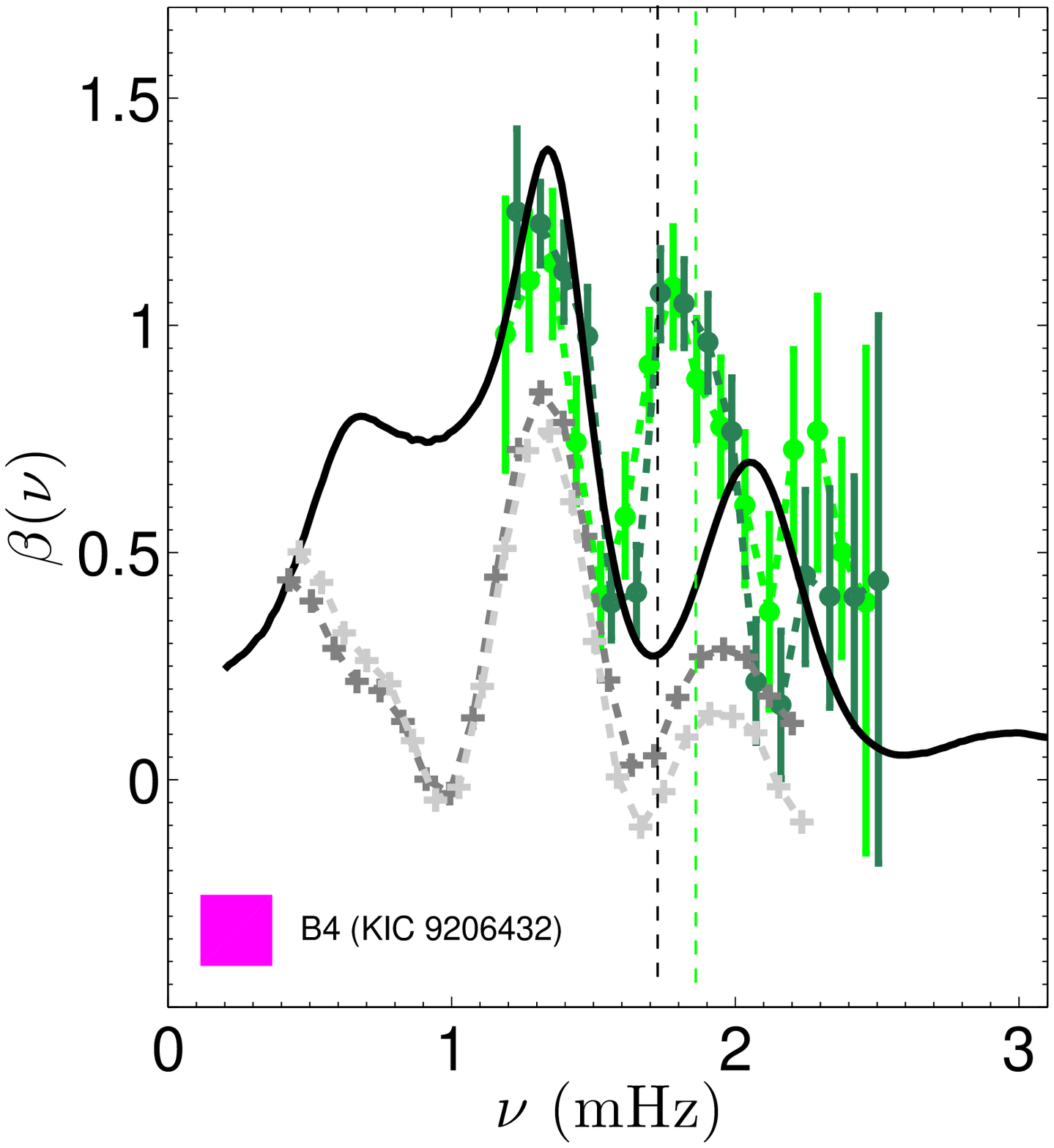}
	\caption{Group B. Black solid line: seismic parameter $\beta(\nu)$ computed from the theoretical model envelope. Gray symbols: theoretical $\beta(\nu)$ computed from a table of frequencies for the degrees $l=0$ (dark gray) and $l=1$ (light gray). Green dashed lines: seismic observable $\beta(\nu)$ computed from an observational frequency table for modes with degree $l = 0$ (light green) and $l=1$ (dark green). The vertical green dashed line indicates the approximate position of the observational frequency at maximum power $\nu_{\text{max}}$. The vertical black dashed line indicates the approximate position of the frequency at maximum power, $\nu_{\text{max}}$, obtained from the scaling relation in \citet{1995A&A...293...87K}. \label{fig:fig5}}
\end{figure*}

	\subsection{The Phase Shift from a Stellar Envelope}

		\subsubsection{The Acoustic Potential}

\begin{figure*}
	\plotone{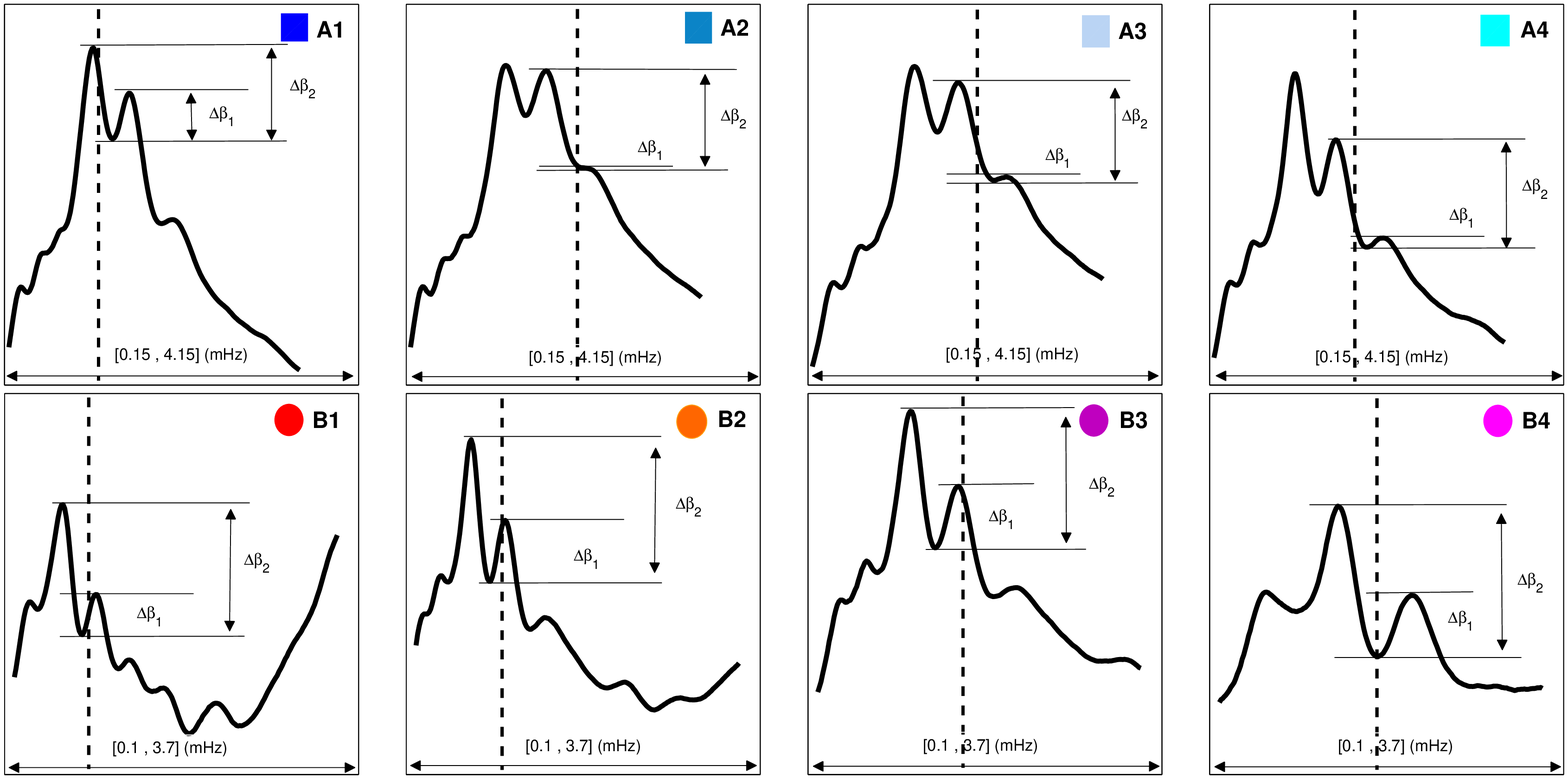}
	\caption{Theoretical seismic signature $\beta(\omega)$ computed from the envelopes of the selected stars. $\Delta\beta_1$ and $\Delta\beta_2$ are the heights of the relative maximums closer to the location of the theoretical value of the frequency of maximum power (from \citealt{1995A&A...293...87K}). The location of   $\nu_{\text{max}}$ is represented by a black vertical dashed line. \label{fig:fig6}}
\end{figure*}

On the surface, and because the contribution of gravity can be neglected (the Cowling approximation), the complete fourth-order system of differential equations describing the linear adiabatic acoustic oscillations can be reduced to a Schr\"{o}dinger-type equation with a suitable choice of variables \citep[e.g.,][]{1979nos..book.....U, 1989ASPRv...7....1V, 2001MNRAS.322..473L}
\begin{equation} \label{eq:(2)}
\frac{d^2 \psi}{d\tau^2} + (\omega^2 - U^2)\psi = 0
\end{equation} 
where $\tau$ represents the acoustic depth, defined as
\begin{equation} \label{eq:(3)}
\tau = \int_r^R \frac{d r}{c} 
\end{equation}
and
\begin{equation} \label{eq:(4)}
\psi = r\sqrt{\rho c} \, \xi
\end{equation}
with $\xi$ being the displacement in the radial direction.
$U^2$ is the reflecting acoustic acoustic potential of the star and is given by
\begin{equation} \label{eq:(5)}
U^2 = \frac{g}{c} \left( \frac{g}{c} - \frac{d\ln h}{d\tau} \right) + \left[ \frac{1}{2} \frac{d \ln \zeta}{ d \tau} \right]^2 - \frac{1}{2}\frac{d^2 \ln \zeta }{d\tau^2}
\end{equation}
with 
\begin{equation} \label{eq:(6)}
\zeta = \frac{r^2 h}{c}
\end{equation}
and 
\begin{equation} \label{eq:(7)}
h(r) = \rho^{-1} \exp \left( -2 \int_0^r \frac{g}{c^2} dr \right) \; .
\end{equation}
In the above expressions, $g$ is the gravitational acceleration, $c$ represents the adiabatic sound speed, and $r$ is the stellar radius. 

A polytropic model of a given index ($n=3$) can be used as an approximation of the background of the solar and stellar envelopes. It revealed itself to be a suitable reference against which to compare a realistic theoretical model. In such a polytropic approximation the acoustic potential is inversely proportional to the square of the acoustic depth, $U^2 \propto 1/\tau^2${\bf{,}} and this stratification was used by \citet{1997ApJ...480..794L} to determine the contributions of the acoustic potential to the seismic parameter $\beta(\omega)$ in the Sun. 
However, deviations from a polytropic stratification induce deviations in the profile of the sound speed and this motivates the analysis of the scattering processes occurring in the external layers of the stars to infer their structure.

The solar acoustic potential has been computed and discussed many times in the past by several authors \citep{1987SvAL...13..179B, 1990SvAL...16..108B, 2001MNRAS.322..473L}. The shape of the solar acoustic potential allows the identification of the superadiabatic region ($\tau \sim 80$ s), the zone of partial ionization of helium at $\tau \sim 600$ s,  and the base of the convective zone at $\tau \sim 2100$ s \citep{1987SvAL...13..179B, 2001MNRAS.322..473L}.

		\subsubsection{The Phase $\alpha$ and Its Derivative: The Seismic Diagnostic $\beta$}

Equation \ref{eq:(2)} can be conveniently solved by the phase-function method \citep{1987SvAL...13..179B, 1989SvAL...15...27B, 1994A&A...290..845L, 1997ApJ...480..794L, 2001MNRAS.322..473L}. In this method the eigenfunction $\psi$ takes the form

\begin{equation} \label{eq:(8)}
\psi(\tau, \omega) = A (\tau, \omega) \cos \left(\frac{\pi}{4} + \pi \alpha(\omega, \tau)  - \omega \tau \right) ,
\end{equation}
and for each value of the frequency $\omega$, the dependence $\alpha(\omega)$ is determined as an asymptotic solution of the Cauchy problem for the phase equation

\begin{equation} \label{eq:(9)}
\frac{d (\pi \alpha)}{d \tau} = \frac{U^2}{\omega} \cos^2 \left( \omega \tau - \frac{\pi}{4} - \pi \alpha(\omega, \tau) \right)
\end{equation}
with sufficiently large values of $\tau$. The boundary condition is imposed at $r = R$ and is taken to coincide with the location of the temperature minimum \citep{1997ApJ...480..794L, 2001MNRAS.322..473L}.
Then, the phase $\alpha$ can  be computed numerically from $\tau = 0$ to a $\tau_{\text{max}}$, where the contribution of the variation of the background state becomes negligible. Usually, for the Sun and solar-type stars, $\tau_{\text{max}}$ is taken near the base of the convective zone.

For a given stellar model $\beta(\omega)$ can be easily calculated from the corresponding acoustic potential. Indeed, following \citet{1987SvAL...13..179B}, the differentiation of equation \ref{eq:(1)} will yield
\begin{equation} \label{eq:(10)}
\beta(\omega) = - \omega^2 \frac{d}{d\omega} \left( \frac{\alpha}{\omega} \right) .
\end{equation}
This function is sensitive to the outermost stellar layers. The contribution of the helium ionization zone, in the solar case, is known to produce a significant periodic component in $\beta (\omega)$ \citep{ 1994A&A...290..845L, 2001MNRAS.322..473L, 2001MNRAS.322...85R}.

The simple procedure we just described to calculate the dependences $\alpha(\omega)$ and $\beta(\omega)$ for a theoretical stellar model envelope offers an effective method for seismically probing the structure of the lower stellar atmospheres and the outer layers of the convection zone in solar-type stars.

\subsection{The Phase Shift from an Acoustic Oscillation Table of Frequencies: The Seismic Observable $\beta$}

By fitting together the solutions obtained for the inner and outer regions of a star, we are establishing a relation between the phase shift frequency dependence and the structure of the outer reflecting layers.
A dispersion relation like equation \ref{eq:(1)} gives the possibility of solving an inverse problem. From a table of frequencies $(\omega, n, \ell)$ it is possible to determine the acoustic phase $\beta(\omega)$. The process of taking the partial derivatives of equation \ref{eq:(1)} with respect to $n$ and $\ell$ yields not only equation \ref{eq:(10)} but also  

\begin{equation} \label{eq:(11}
\beta(\omega) = \frac{\omega - n \left( \frac{ \partial \omega}{\partial n} \right) - L \left( \frac{\partial \omega}{\partial L} \right) }         {\frac{\partial \omega}{\partial n}} ,
\end{equation}
with $L = \ell + \frac{1}{2}$.
This equation offers the possibility of recovering from a given table of frequencies (an eigenvalue matrix $\{ \omega_{\ell, n} \}$) the frequency dependence of the effective phase shift, which will carry valuable information on the physical structure of the stellar interiors.

A virtue of this approach, based on an analysis of the phase shift of the scattering of the acoustic waves in the external regions of stars, is that it offers direct contact with the physical processes occurring in these layers. This is done through the acoustic potential of the star. Changes in the acoustic potential are reflected and magnified in the phase shift and in its derivative.


\section  {Probing the outer layers of Kepler F-stars} \label{sec:3}

\subsection{The Models}

We selected a total of eight Kepler stars of spectral class F to conduct our theoretical analysis of the outer layers. We searched for stars with 10 or more frequencies per degree $\ell$, since the seismic diagnostic $\beta (\omega)$ benefits from a high number of frequencies being used in its computation. Only modes that were correctly detected and fitted were used.
We performed a modeling of the eight selected stars using the evolutionary code CESAM \citep{1997A&AS..124..597M}. This version of CESAM uses the solar mixture of \citet{2009ARA&A..47..481A}, OPAL 2001, as equation of state, and it uses the most recent OPAL opacities. The nuclear reaction rates were computed using the NACRE compilation \citep{1999NuPhA.656....3A}. Eddington's gray law was used to describe the atmosphere and convection was described by the mixing-length theory from \citet{1958ZA.....46..108B} without overshoot. We also included the effects of diffusion according to the Burgers formalism \citep{1969fecg.book.....B}. For each star we compute a grid of models by varying the mixing-length parameter, the mass of the star, and the metallicity  in ranges with physical meaning. For each set of parameters, an automatic search was performed to determine the age that simultaneously reproduces the values of the non-seismic constraints listed in table \ref{table:1}. The spectral type of each star was taken from the astronomical database SIMBAD \citep{2000A&AS..143....9W} whereas the the seismic and non-seismic constraints were taken from 
\citet{2012MNRAS.423..122B, 2012ApJ...749..152M, 2012ApJ...757...99S, 2013ApJ...767..127H, 2013A&A...550A.126M, 2014A&A...571A..35B, 2014A&A...572A..95M}, and \citet{2017ApJ...835..172L}.
The age reached also reproduces the value of the mean large frequency separation that is inferred from scaling relations. Finally, the age is fine-tuned through a minimization of a reduced $\chi^2$ to reproduce the observations as closely as possible, a method identical to that used by \citet{2013ApJ...765L..21C}. The theoretical mode frequencies were computed using the ADIPLS code \citep{2008Ap&SS.316..113C}. The properties obtained for the models of our sample of eight Kepler stars are listed in table \ref{table:2}. The models differ by mass, radius, luminosity, age, and chemical composition, highlighting all the variety of stellar structures that occur even when confined to one spectral class. 

The left panel of figure \ref{fig:fig1} shows an H--R diagram of our models where the luminosity is plotted against the effective temperature. The right panel of figure \ref{fig:fig1} shows the variation with the cyclic frequency ($\omega=2\pi\nu$) of the outer phase shift $\alpha(\nu)$ for these stellar envelopes. This dependency was obtained numerically from equation \ref{eq:(9)} for each of the stellar envelopes represented in table \ref{table:2}.
Figure \ref{fig:fig1} exhibits in an unequivocal manner the relation between $\alpha(\nu)$ and $T_{\text{eff}}$. This is a natural relationship, since the phase shift is determined by the turning points of the acoustic waves and is hence linked with the stellar atmospheres where the effective temperature plays a significant role. Models in figure \ref{fig:fig1} also show a well-known theoretical trend: stars with higher masses are more likely to have lower values of the phase shift values \citep{2011ApJ...743..161W}. 
The dependence between the phase $\alpha$ and the effective temperature was analyzed by \citet{2011ApJ...743..161W, 2011ApJ...742L...3W, 2012ApJ...751L..36W}.

\subsection{The Diversity of Acoustic Cavities Reflecting the Multiplicity of Ionization Profiles}

\begin{figure*}
	\plotone{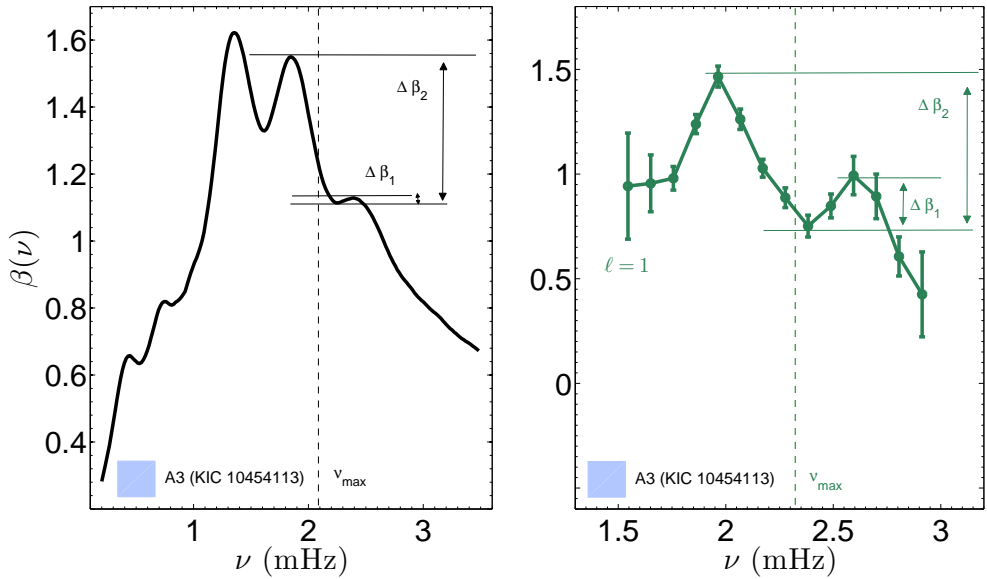}
	\caption{Left: theoretical indexes $\Delta\beta_1$ and $\Delta\beta_2$ for the envelope of the star A3. Right: observational indexes $\Delta\beta_1$ and $\Delta\beta_2$ for the star A3 in the case of the modes with degree $l=1$. The vertical dashed lines signal the locations of the values of the frequency of maximum power, $\nu_{\text{max}}$. In the left panel $\nu_{\text{max}}$ is obtained from scaling relations in \citet{1995A&A...293...87K} and in the right panel it is represented by the observed value of $\nu_{\text{max}}$.   \label{fig:fig7}}
\end{figure*}

\begin{figure*}
	\plotone{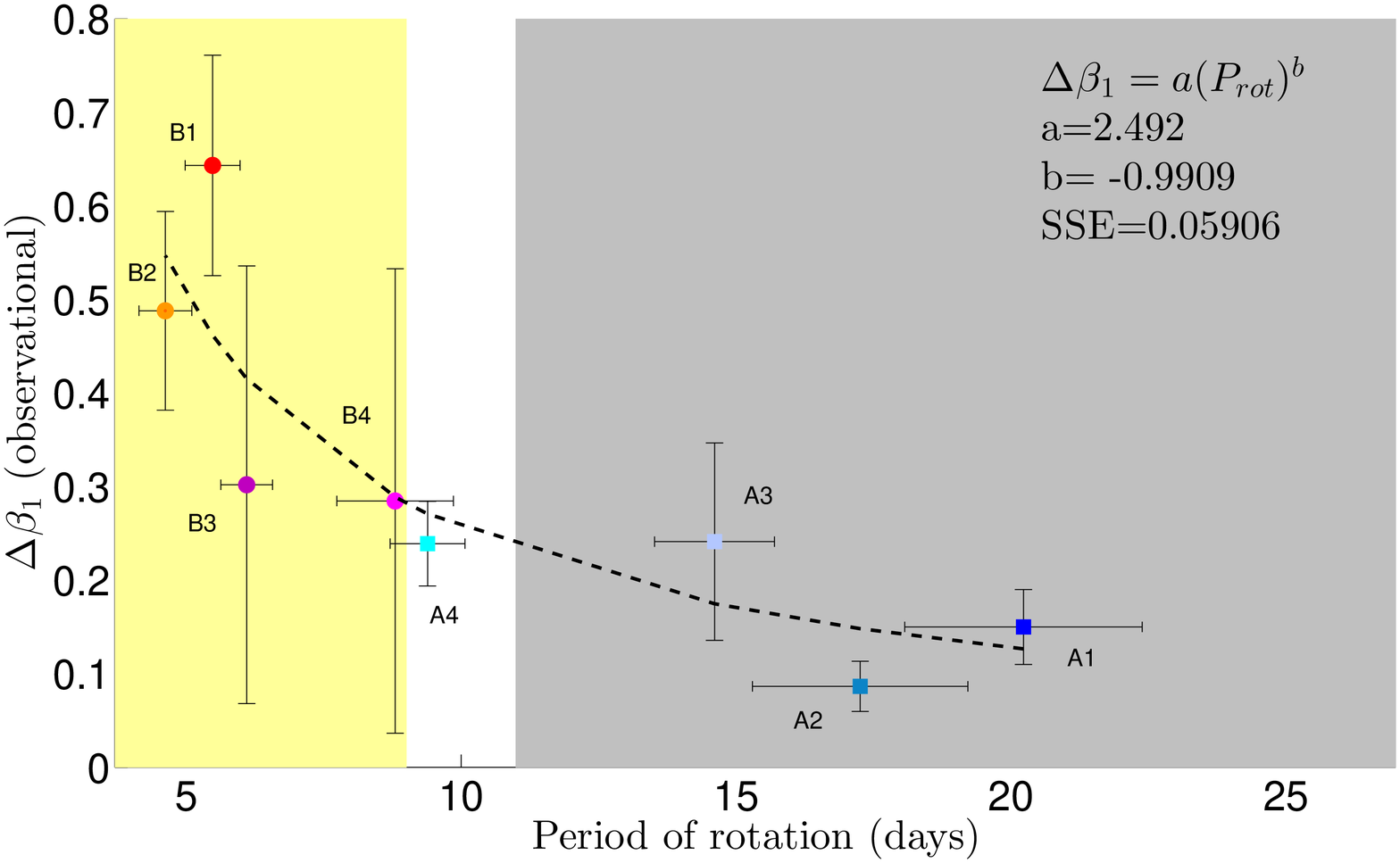}
	\plotone{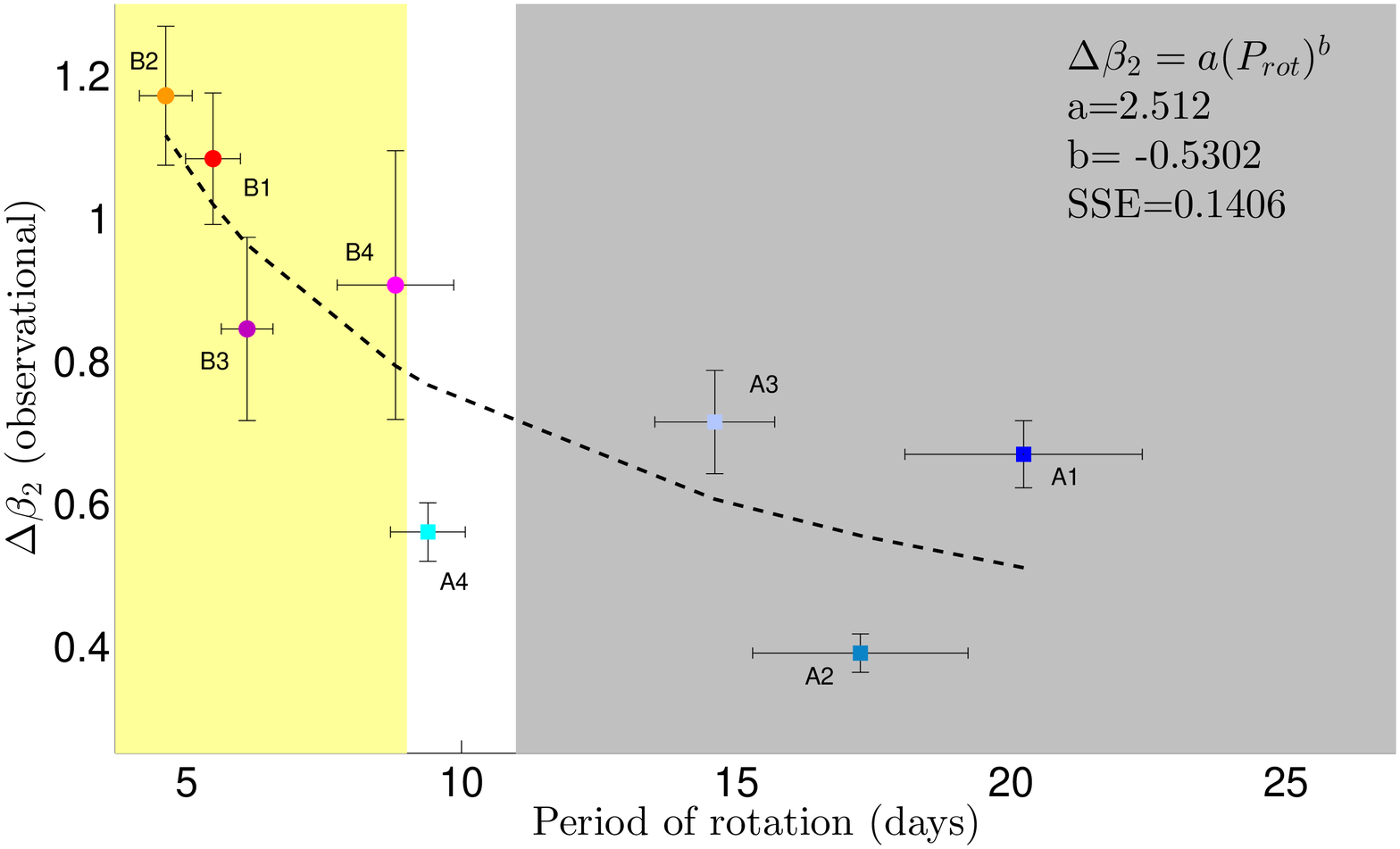}
	\caption{Observational indexes $\Delta \beta_1$ and $\Delta \beta_2$ plotted against the rotation period of the star. To model the relation between the ionization indexes and the stellar rotation, we performed a nonlinear regression using a least squares fit. The dashed black lines, $\Delta\beta_i=a(P_{rot})^b$ ($i=1,2$), represent the results of the fits. Coefficients $a$ and $b$ are given in each figure, as well as the SSE parameter. The stars are identified by the usual colors and abbreviations. The vertical white bar centered at a rotational period of 10 days represents the sharp transition between cool stars rotating slowly (gray region) and hot rapidly rotating stars (yellow region): the Kraft break. \label{fig:fig8}}
\end{figure*}

The acoustic potentials (equation \ref{eq:(5)}) and the corresponding $\beta(\nu)$ signatures for all the eight envelopes of the theoretical models are shown in figure \ref{fig:fig2}. The potentials have been drawn from $r=R$, the location of the temperature minimum, down to the region just before the base of the convective zone in each case. The main difference between the acoustic potentials relates to the partial ionization zones of chemical elements. One clearly notices the diversity associated with the region of the second ionization of helium. For the models of group B, the strong depression that characterizes the region of this ionization is much more pronounced than that in the group A models. Another dip that is possibly associated with the region of the first ionization of helium is clearly visible for the potentials of group B, but is virtually imperceptible for the potentials of group A. The characteristic sinusoidal behavior of $\beta(\nu)$ is related to the region of partial ionizations in the adiabatic convective zone. 
The effect of partial ionizations also leaves a characteristic signature in the first adiabatic exponent $\Gamma_1$ \citep[e.g.][]{2004MNRAS.350..277B}. This signature takes the form of a well-known distinct bump. We display the first adiabatic index for each stellar model in figure \ref{fig:fig3}. 
Figure \ref{fig:fig2} and figure \ref{fig:fig3} reflect the diversity associated with the partial ionization processes in the theoretical envelopes of the solar-type stars. This diversity relates to the locations of the ionization zones and also to the intensity of the ionizations themselves.

In figures \ref{fig:fig4} and \ref{fig:fig5} we present a comparison of the theoretical seismic parameter with the seismic observable $\beta(\nu)$. Both are plotted against cyclic frequency $\nu$.  The theoretical seismic parameter $\beta(\nu)$ was computed using two different approaches: (1) it was obtained from the structural parameters of the stellar model envelopes, and also, (2) it was calculated from a theoretical table of frequencies. In this last case, and to not overload the figures, we presented $\beta(\nu)$ only for the modes with degrees $l=0,1$. Superimposed with the theoretical signatures are the observational $\beta(\nu)$. The observational frequencies and the corresponding uncertainties were taken from \citet{2012A&A...543A..54A} and we have considered, in the calculations, modes with degrees from 0 to 2. Computations were performed over a range of frequencies such that, for each star, only modes correctly detected and fitted as described in \citet{2012A&A...543A..54A} are included. Also, we present the seismic observable $\beta(\nu)$ only for modes with degrees $l=0,1$.

Since the observational frequencies are only available for low-degree modes, they are influenced by the effect of the gravitational potential. On the other hand, the dependence of the phase shift on the frequencies when computed from the structural parameters of the theoretical model envelope is obtained in the Cowling approximation, which in turn neglects the perturbation of the gravitational potential. This is reflected in a well-known mismatch between $\beta(\nu)$ obtained from the acoustic potential and $\beta(\nu)$ obtained from a theoretical table of frequencies in the case of low-degree modes \citep{1997ApJ...480..794L}. This mismatch is more clearly seen in certain models, suggesting that the gravitational potential does not influence the oscillation modes in a similar way for different stars.

Comparing the results for $\beta(\nu)$ that were obtained for the theoretical frequencies and for observational frequencies, it is possible to argue that the differences are not of a purely systematic character. This might indicate that the source of the discrepancy may be related to the internal distribution of the sound speed, thus indicating, it has a thermodynamic nature.
The resulting quasi-periodic behavior of the influence of helium partial ionization in some seismic parameters has been discussed by many authors with different methods \citep[e.g.][]{1990LNP...367..283G, 2006A&A...456..611C, 2007MNRAS.375..861H, 2009arXiv0911.5044H, 2014ApJ...782...16B, 2014ApJ...782...18M, 2014ApJ...790..138V}. In the particular case of the seismic parameter $\beta(\nu)$, the sinusoidal component is known to depend on the contribution of the partial ionization processes. From the comparison between theoretical and observational signatures (figures \ref{fig:fig4} and \ref{fig:fig5}), we note that the amplitudes of the oscillatory components of $\beta(\nu)$ are, in all the cases studied, underestimated by the theoretical models. This underestimation is enhanced in the cases treated in group B. 
For the Sun, the amplitudes of the quasi-periodic component can be related to the abundance of helium in the model \citep{1989ASPRv...7....1V, 1991Natur.349...49V}. Therefore, the underestimation of the amplitudes by theoretical models may indicate that the helium contents in theoretical models are far from the real helium contents of the star.
Another possibility is that in the real star this periodic component is increased by ionizations of other chemical elements \citep{2017MNRAS.466.2123B}.


\section  {The Seismic Diagnostic $\beta$ and the Stellar Rotation Period} \label{sec:4}

	\subsection{The Ionization Indexes $\Delta \beta_1$ and $\Delta \beta_2$}

With the purpose of characterizing the sinusoidal behavior of the seismic parameter $\beta(\nu)$, which we know to be a signature of the partial ionization processes taking place in the model envelopes of solar-type stars, we introduce two indexes. These indexes aim to quantify the amplitude of the periodic component in the seismic parameter $\beta(\nu)$. Therefore, the indexes will act as indicators of the magnitude of partial ionization processes in the outer layers of the model of the star. We define them according to the following procedure. First, we locate the relative minimum, and the two relative maximums, closer to the location of the frequency of the maximum power $\nu_{\text{max}}$. The value for $\nu_{\text{max}}$ is obtained from the scaling relation proposed by \citet{1995A&A...293...87K}. Then we measure the heights of these two relative maximums as illustrated in figure \ref{fig:fig6}. These heights define the indexes
\begin{equation}
\Delta\beta_1=\beta(\nu_{max1})- \beta(\nu_{min})
\end{equation}
and
\begin{equation}
\Delta\beta_2=\beta(\nu_{max2})- \beta(\nu_{min}) \; .
\end{equation}
Here, $\beta(\nu_{min})$ represents the value of the relative minimum located closer to $\nu_{\text{max}}$. $\beta(\nu_{max1})$ and $\beta(\nu_{max2})$ are the values of the two relative maximums closer to $\nu_{\text{max}}$.  The indexes $\Delta\beta_1$ and $\Delta\beta_2$ are represented in figure \ref{fig:fig6} for all the signatures $\beta(\nu)$ that were computed from the theoretical model envelopes. In the same way it is possible to define these indexes also for the theoretical $\beta(\nu)$ computed from a table of frequencies.

Moreover, these indexes can also be computed from the observable $\beta(\nu)$. In this case, they can be measured around the observational frequency of maximum power, as it is shown in figure \ref{fig:fig7}. The observational values of $\Delta\beta_1$ and $\Delta\beta_2$ can then be compared to the predictions of the theoretical values of $\Delta\beta_1$ and $\Delta\beta_2$, revealing information about the microphysics of the outer layers of the star.

	\subsection{Magnetic Activity, Rotation and Ionization}

\begin{figure*}
	\plottwo{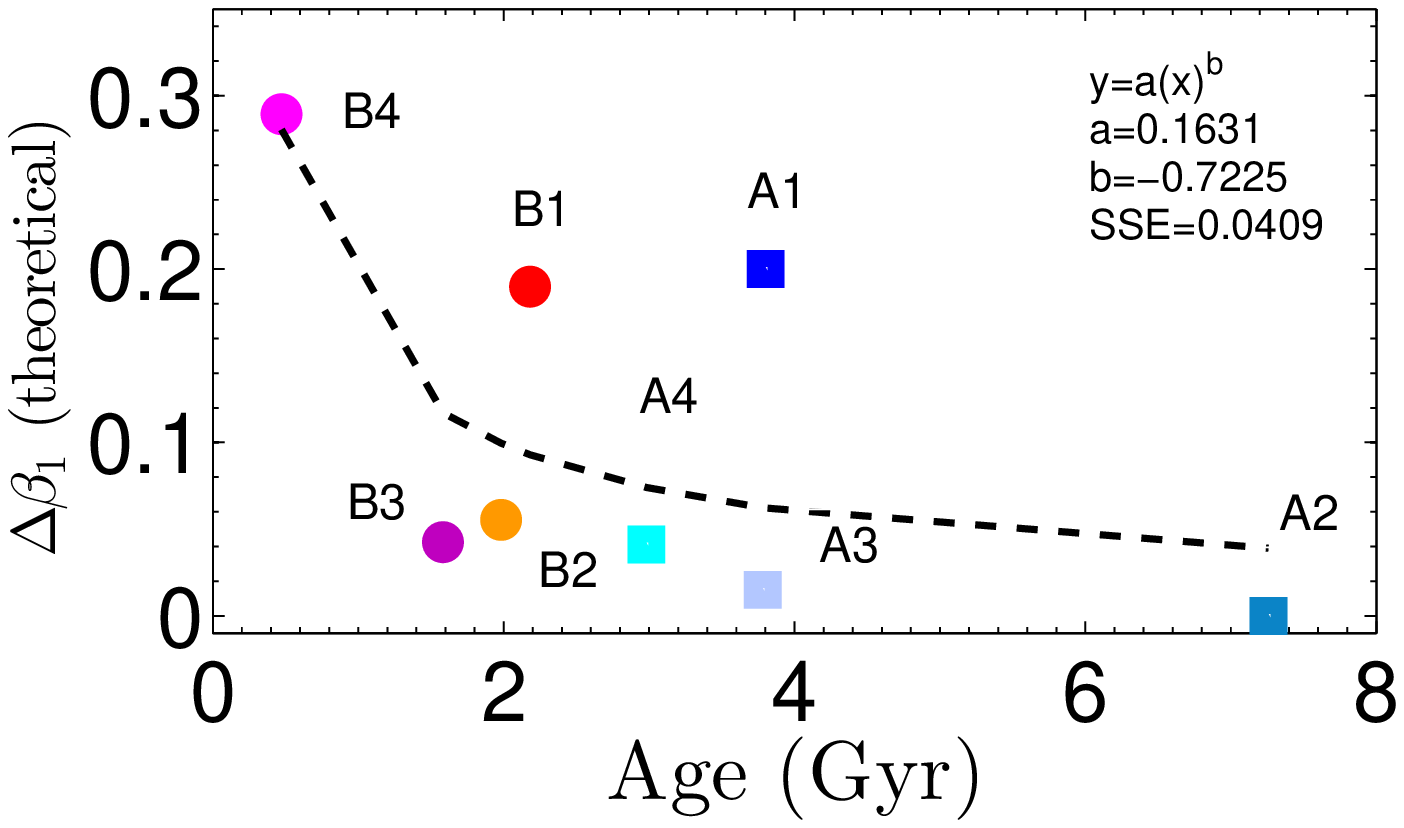}{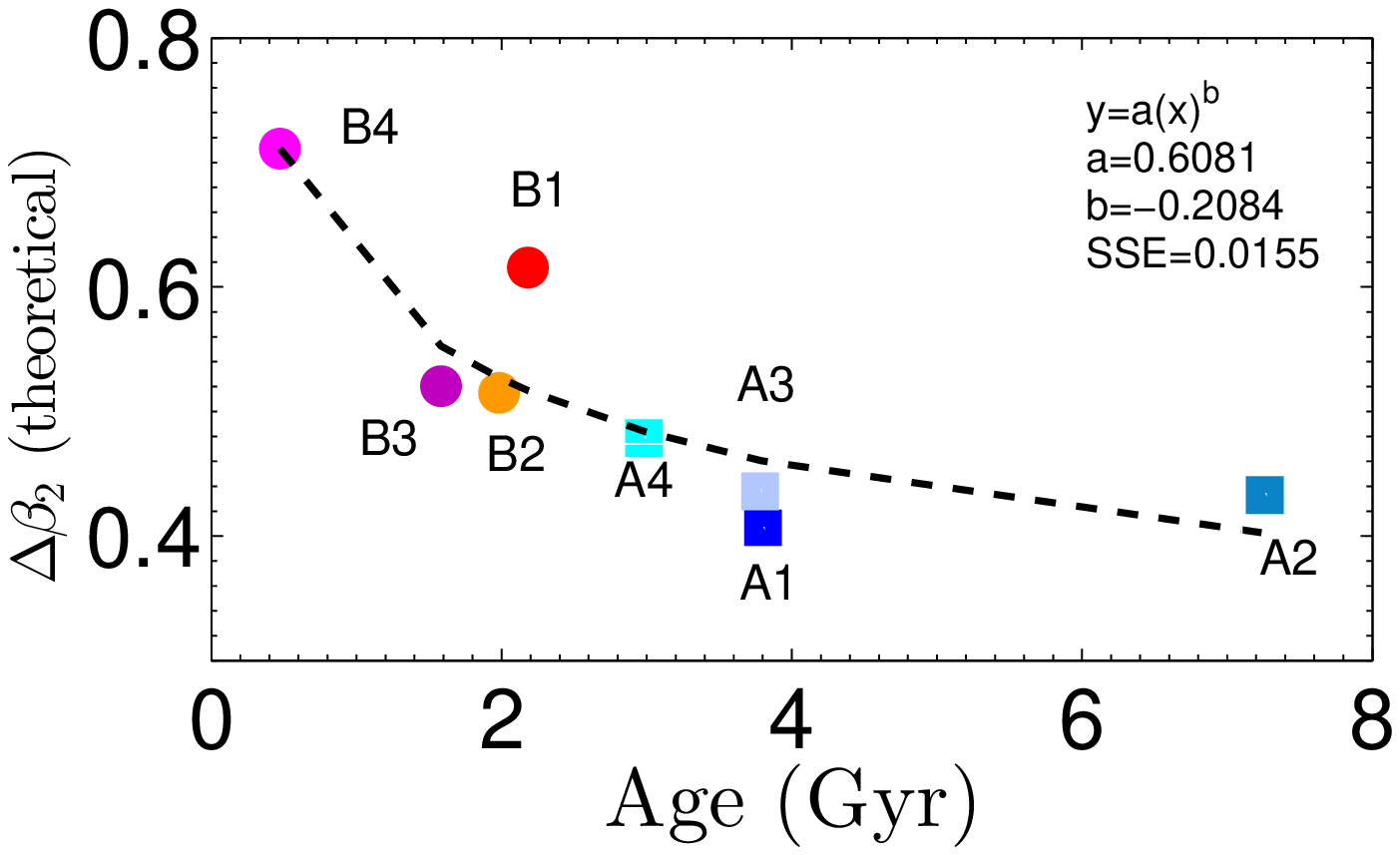}
	\plottwo{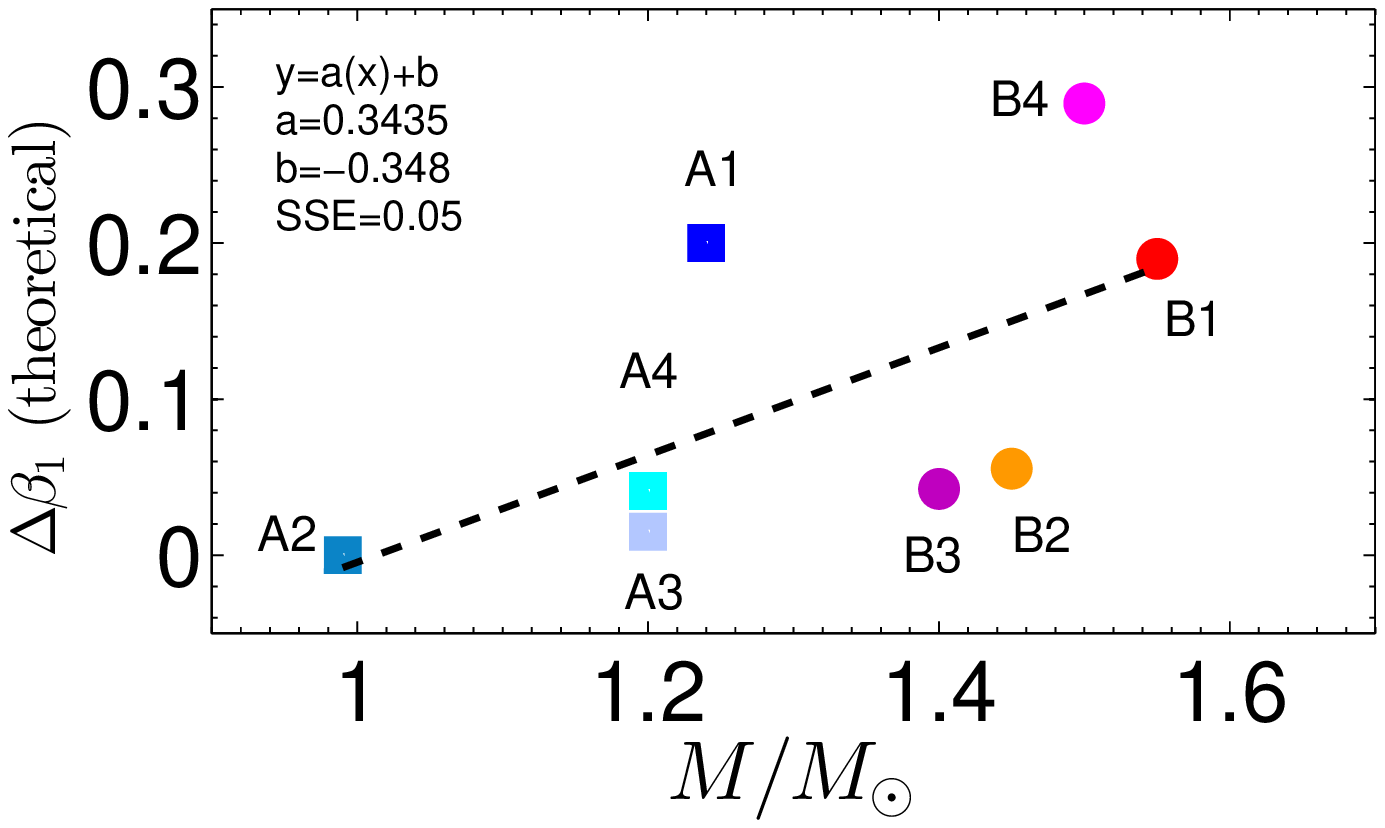}{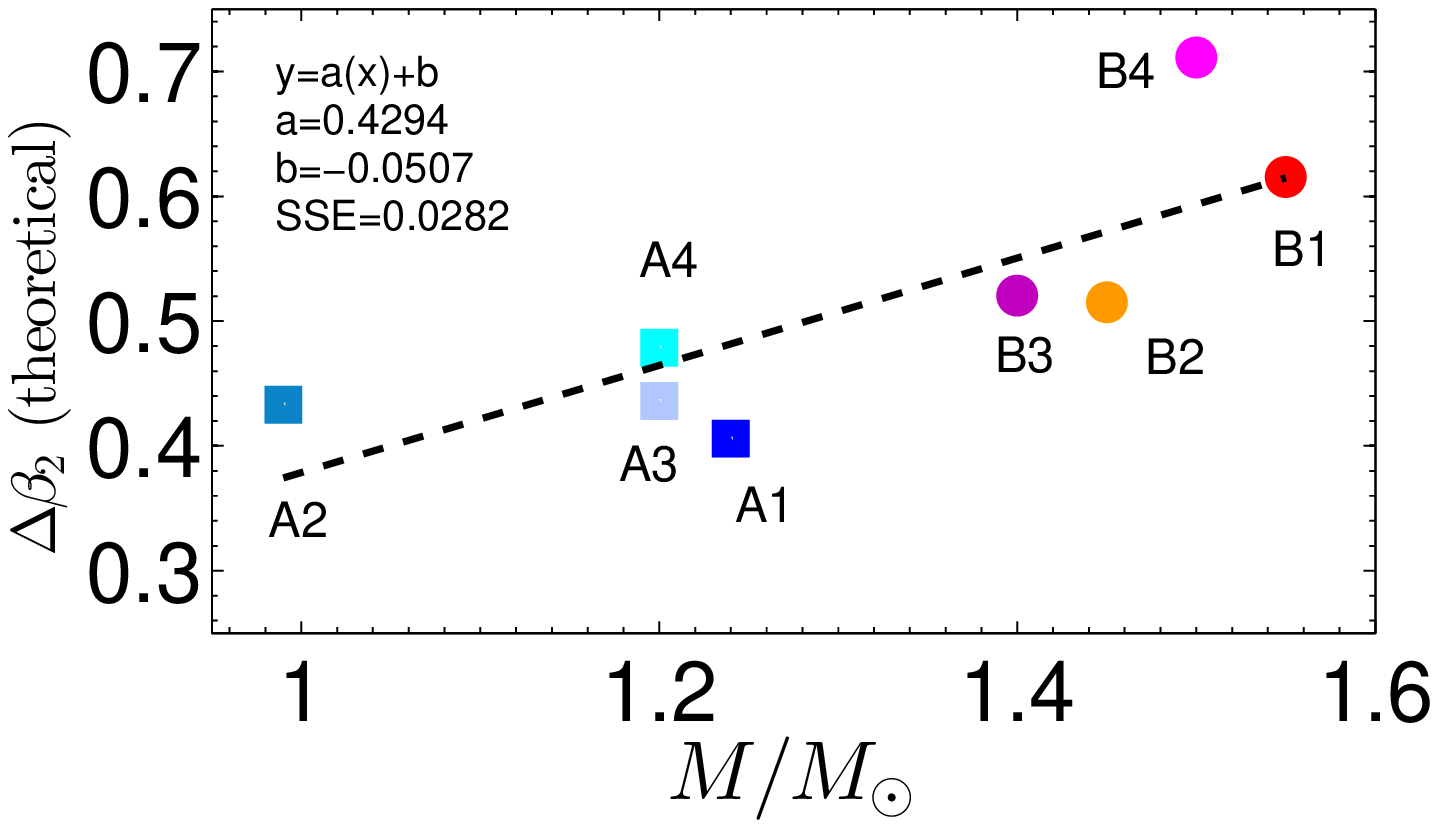}
	\caption{Theoretical indexes $\Delta \beta_1$ and $\Delta \beta_2$ plotted against the values of age and mass of the stellar models. To model the relation between the height $\Delta\beta_i$ ($i=1,2$) and the fundamental parameters of the models, age and mass, we performed regressions using a least squares fit. The dashed black lines represent the results of the fits. Coefficients $a$ and $b$ are given in each figure, as well as the SSE parameter. The stars are identified by the usual colors and abbreviations. \label{fig:fig9}}
\end{figure*}

The influence of magnetic activity on the properties of solar and stellar p-modes is well documented, e.g., \citet{2010Sci...329.1032G}, \citet{2011MNRAS.413.2978B}, and  \citet{2014aste.book....1C}. If magnetic fields can affect the frequencies of the p-modes, they probably can also affect the phase shift of the acoustic waves propagating in this region. Here, we use our results with the phase $\alpha(\nu)$ and the diagnostic $\beta(\nu)$ to learn about a possible influence of magnetic activity on the phases $\alpha(\nu)$ and $\beta(\nu)$. We use the length of the star's rotation period as an indicator of magnetic activity. It is known that rotation and magnetic activity appear to be related. This relation is not always straightforward, and it may even be quite complex and diverse. Fast rotators are generally more active, i.e., they display higher magnetic activity amplitudes than slow rotators. However, fast rotators are less likely to exhibit periodic magnetic behavior than slow rotators. \citep[e.g.][]{1984ApJ...279..763N, 1984ApJ...287..769N, 1996ApJ...457L..99B, 1999ApJ...524..295S, 2002ASPC..277..311S, 2003A&A...397..147P, 2014A&A...562A.124M, 2015SSRv..196..303B, 2015sac..book..437C}. On the other hand, stellar rotation is a crucial property for understanding stellar structure and stellar evolution, since the rotation period of a star is known to be a solid function of the mass and age. 

The two main rotational regimes of stars on the main-sequence were discovered long ago and they differ by the value of the effective temperature of the star \citep{1967ApJ...150..551K}. Cool stars with $T_{\text{eff}}<6200$ K are usually slow rotators. They tend to have rotation periods larger than 10 days. These large rotation periods indicate that the star is loosing angular momentum after entering the main-sequence. The loss of angular momentum is thought to occur due to the action of magnetic winds in the thick convective envelopes of these stars \citep{2009IAUS..258..363I}. The concept of magnetic breaking is used in this context, since the star experiences a spin-down of magnetic origin. This loss of angular momentum is such that stars born with very different rotation periods, after some time in the main-sequence, will all be rotating with similar periods. More precisely, it is possible to say that stars in this regime, at an age of approximately 0.5 Gyr, rotate without the influence of their initial conditions \citep{1989ApJ...338..424P}. It is this behavior of the angular momentum loss that makes gyrochronology possible. Gyrochronology relates the rotation period of a star to the age of the star \citep[e.g.][]{1972ApJ...171..565S, 2007ApJ...669.1167B, 2008ApJ...687.1264M, 2011ApJ...733..115M}. This relation is possible because, according to the description above, all stars of the same age (and mass) will rotate with the same period.
The other rotational regime on the main-sequence is observed in stars with $T_{\text{eff}}>6200$ K. These are hot and rapid rotating stars with thin convective envelopes. The narrowness of the envelopes is thought to be related to the absence of magnetic breaking. This means that, for stars in this regime, the angular momentum loss is minimal and they remain rapidly rotating after entering the main-sequence. If they do not experience a strong spin-down, they are incompatible with gyrochronology \citep[e.g.][]{ 2013ApJ...776...67V, 2014ApJ...780..159E}.

As we just saw, the main aspects of the two rotational regimes in the main-sequence, are intimately connected with processes occurring in the convective envelopes of these stars. At the same time, the convective zones are also the stage for the partial ionizations of the chemical elements that compose the star. Therefore, it seems natural to look for a relation between the specific partial ionization signatures of each star and the rotational period of the star.  Here, we propose to use the ionization indexes $\Delta \beta_1$ and $\Delta \beta_2$ to explore a possible connection between partial ionization processes occurring in the outer layers of solar-type stars and their rotation periods. To this end, we plotted the observational $\Delta \beta_1$ and $\Delta \beta_2$, against the rotation period of the stars.
The results are shown in figure \ref{fig:fig8}. The observational values of $\Delta \beta_1$ and $\Delta \beta_2$ in figure \ref{fig:fig8} were obtained for the modes with the degree $l=1$. The use of $l=1$ modes is due to the fact that they usually have smaller error bars, and at the same time they suffer less from boundary effects during the process of computing numerical derivatives from central differences. Nevertheless, the results still hold for $l=0$ modes, or, even if we consider the mean value obtained for $l=0$ and $l=1$. Modes with degree $l=2$ have usually large error bars, so we did not include them in our study. We found that there is a relation, in the form of a power-law, between the rotation period of a star and the ionization indexes $\Delta \beta_i$ ($i=1,2$).
Rotation periods were taken from \citet{2015MNRAS.452.2654B} and \citet{2017A&A...598A..77K}.
 
The trend obtained with our sample of eight solar-type stars is compatible with the two rotational regimes described above, which are known to be abruptly separated by the Kraft break. This result suggests that partial ionization might be a key process for better understanding rotation in solar-type stars. Moreover, we also plotted  the theoretical ionization indexes $\Delta \beta_1$ and $\Delta \beta_2$ against the age and the mass of our models (figure \ref{fig:fig9}).  The correlations obtained support the previous result since, as is well-known, the rotational period is a function of the mass and age of the star. We think that these ionization indexes may reveal themselves as important diagnostic tools that can improve our knowledge about the structure, and even dynamics, of solar-type stars.

\section{Conclusions}

We studied the outer layers of the theoretical models of eight {\it{Kepler}} F-stars. This was accomplished by using a seismic diagnostic based on the dependence of the surface phase shift of the frequencies, which arises from the reflection of the acoustic modes. Our results allowed us to split the models into two subgroups, A and B, according to the characteristics of the partial ionization zones of their chemical elements. The models of stars in group B show  stronger ionization patterns when compared with the models of stars in  group A.
	
From the point of view of the magnetic activity, F-stars are known to be very interesting, with many of them exhibiting magnetic cycles \citep[e.g.,][]{2014A&A...562A.124M}. Our results seem to be consistent with the paradigm that stars can be divided in two groups according to their level of magnetic activity. Stars represented by the models in the group B have shorter rotation periods, which is an evidence of shorter cycle periods, whereas stars represented by the group A have larger rotation periods and hence longer cycles. The distinction between the theoretical models of groups A and B, which is related to the characteristics of partial ionization processes, seems to suggest that the more ionized a star is, the higher its rotation period. This relation between ionization and rotation can be useful for further understanding the already established relation between rotation and magnetic activity. Indeed, in solar-type stars the magnetic activity is the result of a dynamo process, which in turn results from an interrelation between convection, rotation, and magnetic fields. Therefore, we would expect the presence of a stellar dynamo operating in the real {\it{Kepler}} stars used as targets of this study. The mechanisms of this dynamo should be similar to those in the Sun, with the magnetic field being generated in a thin layer (tachocline) located at the transition region near the base of the convective zone. As the luminosities of the stars increase, the base of the convective zone  gets closer to the surface. For the models in group B, the base of the convective zone is located at a depth of 5-12\% inside the star, whereas for the models in group A, its location is 15-27\% of the radius. This means that, for stars similar to those represented by the group B, the magnetic field will be generated near the surface and very close to the ionization zones.

To help with the description of the partial ionization processes in the theoretical envelopes of these solar-type stars, we introduced two seismic ionization indexes, $\Delta \beta_1$ and $\Delta \beta_2$. They measure the magnitudes of partial ionization processes in the outer layers of the stars. These indexes can be computed from the structural parameters of the theoretical model envelopes, from a theoretical table of frequencies, and also from the observational frequencies. By computing the observational values of $\Delta \beta_1$ and $\Delta \beta_2$, for the sample of the chosen F-stars, we uncovered a relation, in the form of a power law, between these indexes and the rotation period of the star. Stellar rotation is complex since it connects with many unclear aspects of stellar astrophysics. Though, at the same time, it is also a powerful diagnostic tool, since observational data about rotation periods are becoming abundant and of higher quality \citep[e.g.,][]{2013ASPC..479..137N, 2014ApJS..211...24M}. The recent disagreement between asteroseismic data from Kepler and the gyrochronologic relations \citep{2015MNRAS.450.1787A}, and the subsequent finding that the stars that are close to the age of the Sun (and older) are not experiencing the magnetic breaking as they should be, according to the period--age relationships expected by gyrochronology \citep{2016ApJ...826L...2M, 2016Natur.529..181V}, is an example of the complex background that connects rotation, magnetism, and the physics of the interiors of the stars. The results uncovered in this work relating the ionization processes with the rotation period of stars strongly suggest that ionization is an important underlying mechanism for better understanding the relations between microphysics and the physics of rotation and magnetism.

\acknowledgments

We thank to the anonymous referee for comments and suggestions that led to a
more accurate and robust manuscript. We are also grateful to P.
Morel for making available the CESAM code for stellar evolution,
to Jordi Casanellas for the modified version of the same code, and
to J. Christensen-Dalsgaard for his Aarhus adiabatic pulsation code
(ADIPLS). This work was supported by grants from "Funda\c{c}\~ao
para a Ci\^encia e Tecnologia" (SFRH/BD/74463/2010).

%
%

\bibliography{bib_file}

\end{document}